\documentclass{IEEEtran}
\usepackage[dvips]{graphicx}
\graphicspath{{../eps/}{../pdf/}{../jpeg/}}
\DeclareGraphicsExtensions{.eps,.pdf,.jpeg,.png}
\usepackage[cmex10]{amsmath}
\usepackage{xcolor,amsthm, amsfonts, mdwmath, mdwtab, eqparbox, psfrag, setspace}
\usepackage[tight,footnotesize,sf,SF]{subfigure} 
\usepackage{cite,multicol, blindtext,overpic,url,color,tikz,lipsum}
\definecolor{dg}{rgb}{0.1,0.55,0.15}

\usepackage{algpseudocode,algorithm}
\newtheorem{proposition}{Proposition}
\newtheorem{lemma}{Lemma}

\newtheorem{remark}{Remark}

\usepackage{cancel,fancyhdr,acronym}
\usepackage[T1]{fontenc}
\usepackage[normalem]{ulem}
\usepackage{soul}

\newcommand\bout{\bgroup\markoverwith{\textcolor{blue}{\rule[.5ex]{2pt}{1.0pt}}}\ULon}
\definecolor{dg}{rgb}{0.48, 0.25, 0.0}

\usepackage{subfigure}
\usepackage{authblk}

\begin{document}
\setstcolor{blue}
\newlength\mylen
\newcommand\myinput[1]{\settowidth\mylen{\KwIn{}}\setlength\hangindent{\mylen}  \hspace*{\mylen}#1\\}
\setlength{\abovedisplayskip}{4pt}
\setlength{\belowdisplayskip}{4pt}
\setlength{\parskip}{0pt}
\acrodef{bbb}[bbb]{\boldsymbol{\beta^{'}}}
\acrodef{BS}[BS]{base station}
\acrodef{NE}[NE]{Nash equilibrium}
\acrodef{PPP}[PPP]{Poisson point process}
\acrodef{MNO}[MNO]{mobile network operator}
\acrodef{SDC}[SDC]{strict diagonal concave}
\acrodef{MPP}[MPP]{Mat\'ern-hardcore point process}
\acrodef{LHS}[LHS]{left-hand-side}
\acrodef{RHS}[RHS]{right-hand-side}
\acrodef{P.F}[P.F]{proportional fair}
\acrodef{PH}[PH]{Poincar\'{e}-Hopf}
\acrodef{BR}[BR]{Best-reply}
\acrodef{JP}[JP]{Jacobi-play}
\acrodef{CoPSS}[CoPSS]{co-primary spectrum sharing}
\acrodef{OP}[OP]{operator}

\title{Co-primary Spectrum Sharing for Inter-operator Device-to-Device Communication}
\author{
\IEEEauthorblockN{Byungjin Cho, Konstantinos Koufos, Riku J\"antti, and Seong-Lyun Kim} 
\thanks{

B. Cho and R. J\"antti are with the Department of Communications and Networking, Aalto University, Finland (e-mail: byungjin.cho@aalto.fi, riku.jantti@aalto.fi). 

K. Koufos is with the School of Mathematics, University of Bristol, United Kingdom (e-mail: k.koufos@bristol.ac.uk). 

S.-L. Kim is with the School of Electrical and Electronic Engineering, Yonsei University, Korea (e-mail: slkim@yonsei.ac.kr).}}

\maketitle
\begin{abstract}
The business potential of device-to-device (D2D) communication including public safety and vehicular communications will be realized only if direct communication between devices subscribed to different mobile operators (OPs) is supported. One possible way to implement inter-operator D2D communication may use the licensed spectrum of the OPs, i.e., OPs agree to share spectrum in a co-primary manner, and inter-operator D2D communication is allocated over spectral resources contributed from both parties. In this paper, we consider a spectrum sharing scenario where a number of OPs construct a spectrum pool dedicated to support inter-operator D2D communication. OPs negotiate in the form of a non-cooperative game about how much spectrum each OP contributes to the spectrum pool. OPs submit proposals to each other in parallel until a consensus is reached. When every OP has a concave utility function on the box-constrained region, we identify the conditions guaranteeing the existence of a unique equilibrium point. We show that the iterative algorithm based on the OP's best response might not converge to the equilibrium point due to myopically overreacting to the response of the other OPs, while the Jacobi-play strategy update algorithm can converge with an appropriate selection of update parameter. Using the Jacobi-play update algorithm, we illustrate that asymmetric OPs contribute an unequal amount of resources to the spectrum pool; However all participating OPs may experience significant performance gains compared to the scheme without spectrum sharing.
\end{abstract}
\section{Introduction}
\label{sec:Introduction}
The 5th Generation (5G) wireless networks are expected to be much more densely deployed than today's networks due to the rapid increase in the number of connected devices and traffic volumes \cite{Cisco2015}. It is expected to require a 1000 times higher traffic capacity and a 10 to 100 times higher typical user rate \cite{METISD1}. Possible ways to satisfy these increasing demands are to allocate more spectrum and improve spectral efficiency. Since spectrum is rather scarce, especially below 6 GHz \cite{FCC}, mobile network operators, which we will refer to as  \acp{OP} hereafter, will need schemes that utilize spectrum more efficiently. One way to do so is to enable inter-operator spectrum sharing in a co-primary manner \cite{Nokia2}, called \ac{CoPSS}, where multiple \acp{OP} jointly use a part of their licensed spectrum to enable an \ac{OP} to cope with temporary peaks in capacity demand.

Conventionally, operator spectrum allocation has been done in an exclusive manner. Exclusive licensing has well-known advantages including good interference management and guarantee of Quality-of-Service (QoS) for market players, necessary for creating an adequate investment and innovation environment. However, it also suffers from low flexibility and as a result low spectrum utilization might occur. To overcome these limitations, a combination of exclusive spectrum allocation and shared spectrum access has been proposed for 5G systems \cite{METISD1}.

Device-to-device (D2D) communication allows two devices to establish direct communication bypassing the \ac{BS}. \ac{CoPSS} can be used for inter-operator D2D communication, when two users having subscriptions with different \acp{OP} want to communicate directly, and the communication should take place over the licensed spectrum of the \acp{OP}. The only available studies for inter-operator D2D can be found in \cite{Phan2011, Pais2014, Jokinen2014, Cho2015}. The patents \cite{Phan2011, Pais2014} design D2D discovery protocols considering different \acp{OP}. An algorithm for inter-operator D2D spectrum allocation is proposed in \cite{Cho2015}, and an inter-operator D2D trial is presented in \cite{Jokinen2014}, however, the works in \cite{Phan2011, Pais2014, Jokinen2014} do not discuss how to negotiate the amount of spectrum every \ac{OP} is willing to contribute, and the work in \cite{Cho2015} is limited to spectrum negotiations between two \acp{OP} only.

The 3rd Generation Partnership Project (3GPP) is in the process of standardizing D2D communication for 5G networks \cite{3GPP2012, 3GPP2013, 3GPP2014, 3GPP2014_Release13, 3GPP_ran13}. Valuable D2D services provided by the 3GPP system have been identified in \cite{3GPP2012, 3GPP2013, 3GPP2014}, i.e., commercial services and public safety. In \cite{3GPP2014_Release13}, operational requirements for D2D communication are reported, particularly for spectrum operations. The radio resource for two D2D users registered to a Public Land Mobile Network (PLMN) would be under 3GPP network control. The communication of two D2D users registered to different PLMNs is subject to the available spectrum, i.e., the shared Radio Access Network (RAN) \cite{3GPP2014_Release13}.

The sharing of RAN where multiple \acp{OP} share network resources has been established in 3GPP \cite{3GPP_ran13}. The \acp{OP} do not only share the radio network infrastructures but may also share the spectrum, i.e., active RAN sharing. In \cite{3GPP_ran13}, feasible scenarios of inter-operator radio resource sharing are illustrated, i.e., flexibly and dynamically allocating resources on-demand. Since \acp{OP} may have different demands, one requirement is to allocate a different amount of resource to different \acp{OP}, i.e., allocating a certain amount of resource for a specified period of time and/or cells on-demand, or guaranteeing/limiting a minimum/maximum spectrum allocation. However, \cite{3GPP_ran13} does not propose any algorithm to determine the amount of spectrum allocated to each \ac{OP} and address the requirement on inter-operator spectrum sharing for D2D communication.

In parallel with the standardization effort, research is being undertaken to address the fundamental problems in supporting co-primary inter-operator spectrum sharing. In many studies, inter-operator spectrum sharing has been treated as a game where \acp{OP} participating in the game are players, each has an individual utility to maximize, can either cooperate or compete to deal with the strategic interactions of one another for a game-theoretic problem. A cooperative game approach is proposed in \cite{Jorswieck2014, Suris2007, Teng2014} where participating \acp{OP} can obtain the benefits, i.e., fair and efficient spectrum allocation, from exchanging operator-specific information, i.e., channel state \cite{Jorswieck2014}, utility function \cite{Suris2007}, interference price \cite{Teng2014}. However, \acp{OP} are essentially competitors, and may not want to reveal proprietary information to the competitors and other parties. Considering the selfishness of \acp{OP}, spectrum sharing based on a non-cooperative game approach is more appropriate to model and analyze strategic interactions in a co-primary manner.

A non-cooperative game approach has been studied for spectrum sharing between co-primary users \cite{Hailu2014, Lin2009, Gandhi2007, Xu2010, Kamal2009}. In \cite{Hailu2014}, spectrum sharing among co-located RANs based on a one-shot game has been studied, but the equilibrium point under load asymmetry could be inefficient for some \acp{OP} who do not impose any operator-specific constraints on the decision space. In \cite{Lin2009}, a distributed learning algorithm for spectrum sharing has been proposed to increase the convergence rate. The learning rate chosen by each user would result in different convergence rates among the users, thus stability is not ensured. In\cite{Gandhi2007, Xu2010}, auction-based spectrum sharing is studied, where participating \acp{OP} competitively bid for spectrum access through a spectrum broker. In~\cite{Kamal2009}, penalty-based utility functions for spectrum sharing are constructed. Adopting market-driven or punishment-based sharing schemes, however, might not be realistic, because \acp{OP} may not be willing to change their revenue models. Above all, the algorithms \cite{Hailu2014, Lin2009, Gandhi2007, Xu2010, Kamal2009} do not consider the heterogeneity in service type offered by \acp{OP}, i.e., mainly cellular link is considered.

There have been many studies on spectrum sharing for D2D communication, e.g., \cite{Asadi2014} among others, but only single-operator D2D is taken into account. Spectrum allocation for inter-operator D2D based on a non-cooperative game model was first proposed in \cite{Cho2015}. Provided that the game is concave and every \ac{OP} satisfies the diagonal dominance solvability condition (DSC), there exists a unique NE (Nash Equilibrium) and the sequence of best responses converges to it from any initial point. However, the analysis and convergence of the best responses are valid only in a setting with two \acp{OP}. In this paper, we extend the study presented in \cite{Cho2015} by considering \ac{CoPSS} for inter-operator D2D communication with an arbitrary number of \acp{OP}. The main novelty and contributions of this paper can be summarized as follows:
\begin{itemize} 
\item Inter-operator D2D communication over dedicated cellular spectral resources contributed from an arbitrary number of \acp{OP}, is proposed, while intra-operator D2D users subscribed to the same \ac{OP} can use a dedicated resource or reuse the cellular resources of the \ac{OP}. 

\item A general framework for a constraint-based utility maximization problem is proposed, where different preferences of different operators are encompassed. This can be extensively applied to utility and constraint designs under concavity and monotonicity.

\item A non-cooperative game for co-primary interaction is established, where the \acp{OP} make offers about the amount of spectrum committed to the spectrum pool without revealing operator-specific information to other \acp{OP}, and the offers are exchanged by an iterative strategy update algorithm. We show that the Jacobi-play update, with a careful selection of update parameter, can converge to a unique equilibrium with an arbitrary number of \acp{OP}, enabling an \ac{OP} to set the reliable but myopic strategy in a distributed manner.

\item Using the proposed \ac{CoPSS} solution, we show that all \acp{OP} under different intra-operator loads can experience performance gains compared to the baseline scheme where the \acp{OP} do not share spectrum. The efficiency of the non-cooperative game solution is evaluated for different utility designs, weighted sum and weighted proportional fair. 
\end{itemize}

The rest of this paper is organized as follows: In Section II, the system model and problem formulation are presented. In Section III, the non-cooperative game is established with the utility and constraint designs, the existence of a NE, the introduction of the iterative algorithm to reach a NE, and the relation between the uniqueness and local stability properties of a NE. In section IV, the necessary and sufficient conditions for the convergence of the iterative algorithm are provided, and a distributed algorithm always converging to a unique NE is proposed. In Section V, we demonstrate the spectrum sharing gain. Section VII concludes this paper.

\section{\color{black}System model and problem formulation}
\label{sec:System_model}
In this section, we consider the system model and problem formulation used in \cite{Koufos2015}, applicable to increasing the co-primary spectrum usage opportunities for 5G mobile operators.

\subsection{\color{black}System model}
We consider $N$ \acp{OP} enabling D2D communication. An \ac{OP} may have two types of D2D users: firstly, intra-operator D2D (a.k.a. intra-D2D) users, i.e., the two ends in the D2D pair have subscriptions with the same \ac{OP}; secondly, and inter-operator D2D (a.k.a. inter-D2D) users. A D2D pair can communicate in a D2D manner (a.k.a. D2D mode) or via the nearby serving \acp{BS} (a.k.a. cellular mode). D2D communication in unlicensed bands would suffer from unpredictable interference. Because of that, at this moment, licensed spectrum seems to be the way forward to enable D2D communication, especially considering safety-related scenarios such as vehicle-to-vehicle communication. In the licensed band, either dedicated spectrum can be allocated to the D2D users (a.k.a. D2D overlay) or D2D and cellular users can be allocated over the same resources (a.k.a. D2D underlay). An intra-D2D underlay has a higher spectrum reuse factor and it may result in high spectral efficiency with appropriate interference management. An intra-D2D overlay enjoys higher spectral efficiency than the case where D2D communication is not enabled, and less implementation complexity than the underlay case \cite{Asadi2014}. In an inter-D2D underlay, cellular users may suffer from inter-operator interference, and in order to resolve it, information exchange between the \acp{OP} might be needed. Due to the fact that \acp{OP} may not be willing to reveal proprietary information, we believe that, at the first stage, the inter-D2D overlay scheme would be easier to implement \cite{Cho2015}.

Each \ac{OP} utilizes the total available spectrum divided into multiple sub-bands. Without loss of generality, we assume a fixed and equal bandwidth for sub-bands to easily determine the total available sub-bands. The number of available sub-bands are equally allocated among all users for fairness of resource usage, or may be based on a set QoS per user. We assume a fair share of resources among users. In particular, in \ac{CoPSS}, whether or not an \ac{OP} may contribute an equal amount of spectrum to the shared band, the performance accessible to inter-D2D users in D2D mode should be proportional to the bandwidth allocated to the spectrum pool under the fairness rule~\cite{Koufos2015}. In inter-D2D and intra-D2D modes, each user accesses a sub-band with a certain medium access probability. In cellular mode, the resource allocation is controlled by the BS which schedules the users in a round-robin manner, and thus the corresponding spectrum resource is shared equally. Accordingly, the average numbers of users over sub-bands allocated for each transmission mode are the same.

\begin{figure}[t!]
\centering
\begin{tabular}{c} \hspace{-.152in}
\includegraphics [width=0.485\textwidth]{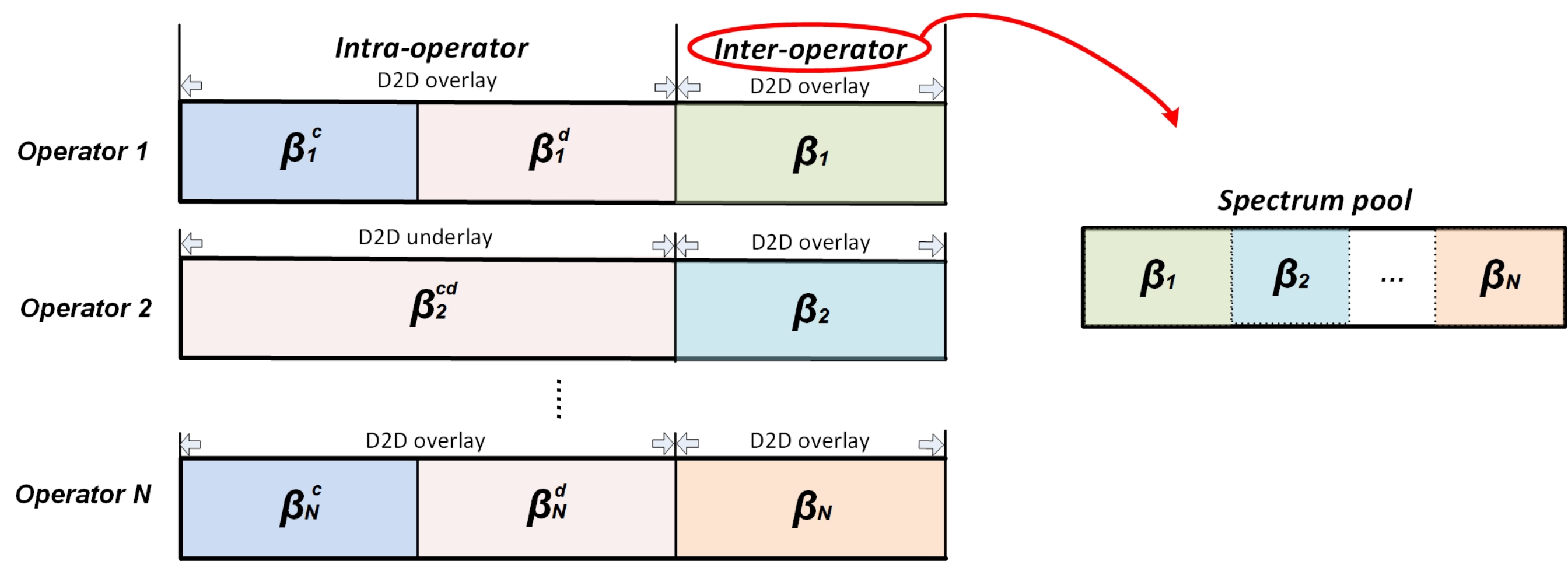} 
\end{tabular}
\caption{Spectrum divisions for inter-operator D2D communication.}
\label{Numerical:fig:scen}
\end{figure}

Fig.~\ref{Numerical:fig:scen} shows the spectrum allocation for the \acp{OP} in case they employ the overlay principle for inter-D2D communication, while intra-D2D communication can be either in overlay or underlay. A fraction $\beta_i^{cd}$ of the $i$-th \ac{OP}'s spectrum is dedicated for cellular and intra-D2D communications. The fraction $\beta_i^{cd}$ is further divided into two sub-fractions in an intra-D2D overlay scheme, $\beta_i^{cd} = \beta_i^c + \beta_i^d$ where $\beta_i^c$ and $\beta_i^d$ are fractions of the $i$-th \ac{OP}'s spectrum, dedicated for cellular and intra-D2D communications, respectively. No matter which scheme is used for intra-D2D communication, an \ac{OP} contributes a fraction $\beta_i$ of spectrum to the spectrum pool, $\beta = \sum_i \beta_i$, where inter-D2D communication takes place. An inter-D2D pair can use any of the resources contributed to the pool. Obviously, $\beta_i^{cd} +\! \beta_i \!=\! 1, \forall i$. While in our analysis we assume FDD \acp{OP} that contribute frequency resources for D2D communication, the same analysis is applicable to TDD \acp{OP} that contribute time-frequency resource blocks with time synchronization among the \acp{OP}.

The mode selection decides whether a D2D pair should be communicating in D2D or in cellular mode. The mode selection algorithm is not necessarily the same for different \acp{OP} neither for inter-D2D and intra-D2D communications. Mode selection determines the density of D2D transmissions, thus the potentials for increasing the frequency reuse factor. At the same time, it affects the amount of interference among the intra-D2D users and the cellular users in a D2D underlay, and the amount of the D2D self-interference in a D2D overlay. As a result, mode selection and spectrum allocation for D2D communication are coupled in the system design. For instance, in dense deployments, a mode selection resulting in high inter-D2D density and thus in high self-interference in D2D can be compensated by allocating more spectrum for inter-D2D communication. This means there would be less spectrum available for cellular and intra-D2D transmissions in an inter-D2D overlay.

In literature, mode selection for single-operator D2D utilizes either D2D pair distance~\cite{Lin2013} and/or the distance between the D2D transmitter and cellular \ac{BS}~\cite{Yu2011}, or energy-based detection threshold \cite{Cho2014} as selection criterion.  In a setting with multiple \acp{OP}, implementation of mode selection is not straightforward: Participating \acp{OP} need to agree about the mode selection criteria in the spectrum pool. Also, they need to decide which \ac{OP} should be responsible for taking the decision and communicating it to the users. In this paper, we investigate the spectrum allocation problem, while the mode selection is not treated in detail. Instead, we model the impact of mode selection through the fraction of intra-D2D users selecting D2D mode, denoted by $\delta_i$, and the fraction of inter-D2D users selecting D2D mode, denoted by $q$. The parameters, $\delta_i$ and $q$, are related. For instance, when the value of $q$ increases, more inter-D2D users would select D2D mode, or equivalently less inter-D2D users would communicate over cellular resources. This would affect the parameter settings, e.g., $\delta_i$ and $\beta_i$, which determine the performance accessible to cellular and intra-D2D users. This model can be extended to incorporate the impact of more general mode selection algorithms on the performance gain for a future study.

\subsection{Problem formulation} \label{sec2_2}
Each \ac{OP} must experience performance gain from enabling inter-D2D communication. Such a gain can be quantified by excess utility showing the difference between the \ac{OP}'s utility, $U_i$, when spectrum sharing is used, i.e., $\!\beta_i\!>0$, and the utility, $U_i^{min}$, when spectrum sharing is not used, i.e., $\!\beta_i\!=\!0$. When the utility of an \ac{OP} with spectrum sharing is lower than the utility with no sharing, $U_i \!<\! U_i^{min}$, the spectrum would not be shared. Inter-operator D2D support poses a requirement for exchanging signaling information between the \acp{OP}. Because of that, when spectrum sharing takes place, we assume that an \ac{OP} contributes at least a small positive fraction of spectrum, $\!\beta_i^{min}\!>\!0$, i.e, $\beta_i\!\geq\!\beta_i^{min}$, for the signaling channel. \acp{OP} agree a priori to use a common decision threshold for selecting the inter-D2D mode. This threshold can be mapped to a fraction of inter-D2D users communicating over D2D, $q$. \acp{OP} are free to optimize the fraction of intra-D2D users communicating over D2D, $\delta_i$. Hence, the value of $\delta_i$ is not necessarily equal to the corresponding fraction without spectrum sharing.

The utility of an \ac{OP}, $U_i$, may depend on the amount of spectrum all \acp{OP} contribute to the spectrum pool, $\beta$, while the utility, $U_i^{min}$, does not. Given the aggregate proposal from the opponents, $\beta_{-i} = \sum_{j\neq i}{\beta_j}$, each \ac{OP} identifies its contribution, $\beta_i$, and intra-D2D mode selection parameter, $\delta_i$, for maximizing the utility, $U_i$, under operator-specific constraints. These constraints can, for instance, refer to the rates of cellular users and intra-D2D users in D2D mode. The constraint functions for cellular users and intra-D2D users in D2D mode could be larger than target values, respectively, $h_i^c\geq \tau_i^c$ and $h_i^d \geq \tau_i^d$. The utility, $U_i^{min}$, and the constraint functions for no spectrum sharing can be evaluated in advance. We assume that feasible target values are selected so that these constraints without spectrum sharing, i.e., $\beta\!=\!0$ and $q=0$, are satisfied. To sum up, the amount of spectrum to be contributed for inter-D2D communication, $\beta_i$, and the fraction of the intra-D2D users in D2D mode, $\delta_i$, could be identified as follows 
\begin{subequations}
\label{eq:P1}
\vspace{-.15in}
\begin{align}
\underset{\begin{subarray}{ll}\beta_i, \delta_i \end{subarray}}{\rm{Maximize:}} & \quad U_i({\beta}_i, \delta_i).   \label{eq:P1:a}\\ 
     {{\rm{Subject}}\,\,{\rm{to:}}} & {~h_i^c(\beta_i, \delta_i) \geq \tau_i^c}. \label{eq:P1:b}\\
      & {~h_i^d(\beta_i, \delta_i) \geq \tau_i^d}. \label{eq:P1:c} 
\end{align}
\end{subequations}

We assume that the utility function, $U_i$, in~\eqref{eq:P1:a} is concave in $\beta_i$ for a fixed $\delta_i$. The concavity of the utility with respect to $\beta_i$ indicates that the marginal utility decreases with a further increase in $\beta_i$; Diminishing marginal utility means that the more an \ac{OP} contributes spectrum to the spectrum pool, the less utility gain is obtained. Once the maximum of $U_i^{}$ is achieved, any further increase in $\beta_i$ may decrease $U_i^{}$. For instance, if $U_i$ incorporates both inter-D2D and cellular or intra-D2D user rates, allocating more spectrum in the spectrum pool could increase the rate of inter-D2D users in D2D mode, but at the same time less spectrum becomes available for cellular and intra-D2D transmissions. We assume that the constraint functions, $h_i^c$ and $h_i^d$ are concave in $\beta_i$ for fixed $\delta_i$. These assumptions allow the one-dimensional constraint set of an \ac{OP} to be convex, closed, and bounded. Thus, if there is always a $\beta_i$ such that the constraints in~\eqref{eq:P1:b} and~\eqref{eq:P1:c} are strictly satisfied, the first-order KKT conditions are both necessary and sufficient.

In an intra-D2D underlay, we assume that constraint functions, $h_i^c$ in~\eqref{eq:P1:b} and $h_i^d$ in~\eqref{eq:P1:c}, are decreasing in $\beta_i$ for a fixed value of $\delta_i$, because the cellular and intra-D2D user rates should be increasing functions of the allocated bandwidth $\beta_i^{cd}$. This would make $h_i^c$ and $h_i^d$ decreasing functions. Let ${\beta_i^{cd}}$ denote the amount of spectrum fraction for cellular and intra-D2D communications, satisfying the constraints \eqref{eq:P1:b} and \eqref{eq:P1:c}. Due to the fact that the \acp{LHS} of \eqref{eq:P1:b} and \eqref{eq:P1:c} are increasing in $\beta_i^{cd}$, the two minimum values of $\beta_i^{cd}$ satisfying the constraints \eqref{eq:P1:b} and \eqref{eq:P1:c} make the \acp{LHS} of \eqref{eq:P1:b} and \eqref{eq:P1:c} equal to $\tau_i^c$ and $\tau_i^d$, respectively. Using $\beta_i^{cd}\! = \!1-\!\beta_i$, the constraints \eqref{eq:P1:b} and~\eqref{eq:P1:c} can be converted to a single inequality constraint, $1\!-\!\beta_i\! \geq\! \beta_{i}^{cd,min} \!\!=\!\! \max(\beta_{i,c}^{cd},\beta_{i,d}^{cd})$ where $\beta_{i,c}^{cd}$ and $\beta_{i,d}^{cd}$ satisfy~\eqref{eq:P1:b} and \eqref{eq:P1:c} with equality, respectively.

In an intra-D2D overlay, we assume that each \ac{OP} has a utility function considering the D2D user rates so that the maximum of the utility occurs along the feasibility border of the spectrum allocation factor $\beta_i^c$ with respect to $\delta_i$. Thus, for a fixed $\delta_i$, the constraint function for cellular users could be equal to a target value, $h_i^c \!=\! \tau_i^c$, and the optimization property is the same as in the intra-D2D underlay case. Let ${\beta_i^{c,min}}$ denote the amount of spectrum fraction for cellular communication, satisfying the constraint~\eqref{eq:P1:b} with equality. Due to the fact that the \ac{LHS} of \eqref{eq:P1:c} is increasing in $\beta_i^d$, the minimum $\beta_i^d$ satisfying the constraint~\eqref{eq:P1:c} makes the \ac{LHS} of \eqref{eq:P1:c} equal to $\tau_i^d$. Using $\beta_i^d \!\!=\!\! 1\!-\!\beta_i^c \!-\!\beta_i$, the constraints~\eqref{eq:P1:b} with equality and \eqref{eq:P1:c} can be converted to an inequality constraint, $1\!-\!\beta_i \!\geq\! \beta_i^{c,min} \!\!+\!\! \beta_i^{d,min}$ where $\beta_i^{d,min}$ satisfies \eqref{eq:P1:c} with equality. As a result, the upper limits of the constraint set, $\beta_i^{max}\!=\!1\!-\!\beta_i^{cd,min}$ in an intra-D2D underlay and $\beta_i^{max} \!=\! 1\! -\! \!\beta_i^{c,min}\! \!-\!\! \beta_i^{d,min}$ in an intra-D2D overlay do not depend on the proposals from the opponents, which result in a box constraint, $\beta_i^{min}\!\leq\! \beta_i\! \leq\!\beta_i^{max}$, respectively, for a fixed $\delta_i$.

While the problem in~\eqref{eq:P1} is concave in $\beta_i$, it is not known to be jointly concave in both $\beta_i$ and $\delta_i$. {In case of the nonconvex problem in~\eqref{eq:P1},} finding an optimal solution may be intractable, since the first order conditions are necessary, but not always sufficient. Note that the optimal solution of the problem in \eqref{eq:P1} exhibits the intuitive monotonicity property. The use of the monotonicity property would substantially alleviate the difficulty in obtaining the optimal solution of the problem in \eqref{eq:P1}, by allowing some solution to be on the boundary of the feasible set. For the monotonic solution, according to Topkis's theorem \cite{Topkis}, the constraint set needs to have an ascending or descending property. This property is satisfied if the boundaries of the constraint set are increasing or decreasing functions of a parameter. We assume that the upper limits of $\beta_i$ are increasing in $\delta_i$, determined by $\beta_i^{c,min}$ and $\beta_i^{d,min}$ in an intra-D2D overlay, and $\beta_i^{cd,min}$ in an intra-D2D underlay, respectively. Thus, if $\beta_i^{c,min}$, $\beta_i^{d,min}$, and $\beta_i^{cd,min}$, are decreasing in $\delta_i$, $\beta_i^{max}$ is increasing in $\delta_i$ and thus the constraint sets hold the ascending property \cite{Topkis}.

In an intra-D2D overlay, we assume that $\!\beta_i^{c,min}\!\!$ and $\beta_i^{d,min}\!\!$ are decreasing in $\delta_i$. The LHSs of \eqref{eq:P1:b} and \eqref{eq:P1:c} are increasing in $\delta_i$ for a fixed value of $\beta_i$, because more intra-D2D users in D2D mode yield more time resources available for cellular-based communication and more concurrent D2D transmissions, resulting in increasing rates of cellular users and intra-D2D users in D2D mode. Due to the increased interference, D2D link rate might decrease but the increasing fraction of intra-D2D users in D2D mode dominates the rate in D2D mode, which will be justified later. This would make $h_i^c$ and $h_i^d$ increasing functions of $\delta_i$. Since $h_i^c$ and $h_i^d$ are also increasing in $\beta_i^{c}$ and $\beta_i^{d}$, respectively, less $\beta_i^{c}$ and $\beta_i^{d}$ could sustain certain values of $h_i^c$ and $h_i^d$ with more $\delta_i$. This would make $\beta_i^{c,min}$ and $\beta_i^{d,min}$ decreasing in $\delta_i$.

In an intra-D2D underlay, we assume that $\!\beta_i^{cd,min}\!\!$ is decreasing in $\delta_i$, where $\beta_i^{cd,min}$ is determined by the two minimum values of $\beta_i^{cd}$, $\beta_{i,c}^{cd}$ and $\beta_{i,d}^{cd}$ satisfying the constraints in \eqref{eq:P1:b} and \eqref{eq:P1:c} with equalities, respectively. While $h_i^d$ in~\eqref{eq:P1:c} is increasing with $\delta_i$ as in an intra-D2D overlay case and thus $\beta_{i,d}^{cd}$ is decreasing in $\delta_i$, $h_i^c$ in~\eqref{eq:P1:b} stays almost constant or even slightly decreases with $\delta_i$ \cite{Lin2013}, due to the interference caused by underlaid D2D transmissions. The selected target value for the D2D mode, $\tau_i^d$, is assumed to be large enough such that it is usually larger than the one for the cellular mode, $\tau_i^c$, $\beta_{i,d}^{cd}$ is larger than $\beta_{i,c}^{cd}$, and the cellular users with $\beta_{i,d}^{cd}$ are able to achieve the performance strictly larger than $\tau_i^c$. Thus, $\beta_{i,d}^{cd}$ determines $\beta_{i}^{cd,min}$ which is decreasing in $\delta_i$.  Thus, the upper limits of $\beta_i$, $\beta_i^{max} = 1-\beta_i^{cd,min}$ in an intra-D2D underlay and $\beta_i^{max} = 1 - \beta_i^{c,min} \!- \beta_i^{d,min}$ in an intra-D2D overlay, are increasing function of $\delta_i$, holding the ascending property.

\section{\color{black}Non-cooperative game model}
We consider a strategic non-cooperative spectrum sharing game among $N$ \acp{OP}, $\mathcal{G} \!=\! (\mathcal{N}, \mathcal{S} , \mathcal{U})$, where $\mathcal{N}$ is the set of \acp{OP}, $\!\mathcal{S}\!\!=\!\! S_1 \!\times\! \cdots \!\times\! S_N$ is the set of the joint strategies, and ${\mathcal{U}} \!=\! [U_1,\!\cdots\!,\!U_N]$ is the vector of utilities. The strategy space for an \ac{OP} represents the spectrum fraction contributed to the spectrum pool, i.e., $S_i\!\!=\!\!\{\beta_i\!:\!\! \beta_i^{min} \!\!\leq\! \beta_i\! \leq \beta_i^{max}(\delta_i)\},\,\forall i$.

\subsection{Design of operator specific utility and constraints} 
Each user calculates it's utility with the information of the performance locally measured or obtained via feedback channel from the receiver, and then sends it to the BS. With the utility information of all users, each \ac{OP} can calculate the average utility and then broadcast it to all users for their intra-D2D mode selection decision which determines the density of active D2D users in the intra-D2D mode, $\delta_i$ and spectrum allocation factor, $\beta_i$. \acp{OP} may have different preferences concerning the spectrum allocation formulated by means of a utility function. Considering the different types of users, a utility in (\ref{eq:P1:a}) can be expressed as $U_i \!=\! u_i(Q_i^c,\!Q_i^d,\!Q_i^s)$ where $Q_i^k$ is the (normalized) average rate of the $k$-th type users, $k\!\in\!\{c,d,s\}$, and $c$, $d$ and $s$ correspond to cellular, intra-D2D and inter-D2D users. $u_i(\cdot)$ can take different forms, e.g., weighted sum rate function: $\!\sum_k \!w_i^k Q_i^k$, or weighted \ac{P.F} function: $\!\sum_k w_i^k \log Q_i^k$ where $w_i^k \geq 0$ are weights indicating the normalized densities of the $k$-th type users.

The average rate, $Q_i^k$, refers to the ability of the $k$-th type users to achieve a certain level of data rate performance, which is generally assessed by scaling their average spectral efficiency, $R_i^m$, with the normalized bandwidth, $\beta_i^m$, available for every $m$-th type link mode, and summing it for all transmission modes, $\forall m\in\{c,d,s\}$ where $c$, $d$ and $s$ correspond to cellular, intra-D2D and inter-D2D transmission modes, i.e., $Q_i^c = \beta_i^c  \, R_i^c$ for cellular users, $Q_i^d = \beta_i^c  \, R_i^c  (1\!-\!\delta_i) \!+\! \beta_i^d  R_i^d \delta_i$ for intra-D2D overlay users, and $Q_i^s = \beta_i^c  \, R_i^c  (1\!-\!q) \!+\! \beta R^s q$ for inter-D2D overlay users where $\delta_i$ and  $q$ are fractions of intra-D2D users and inter-D2D users selecting respective D2D modes. We assume that the constraint functions for cellular users and intra-D2D users in D2D mode are $h_i^c = \beta_i^c R_i^c$  and $h_i^d = \delta_i \beta_i^d R_i^d$, respectively. For intra-D2D underlay mode, $Q_i^c$, $Q_i^d$, $Q_i^s$, $h_i^c$, and $h_i^d$ can be obtained by replacing $\beta_i^c$ and $\beta_i^d$ with $\beta_i^{cd}$.

The average rate of the $m$-th type link mode can be obtained as a function of SINR in semiclosed form by averaging over the distribution of the coverage probability, expressed as $R_i^m \!\!=\!\! E[\nu_i^m \log(1+\gamma_i^m)]\! \!=\!\!  \int_0^{\infty}{\frac{\nu_i^m\mathcal{P}_i^m}{(1+\gamma)} d\gamma}$ where $\gamma$ is the SINR target, $\gamma_i^m$ is the SINR, $\mathcal{P}_i^m$ is the coverage probability and $\nu_i^m$ is the portion of time a typical user is active in the $m$-th type link mode. The index $i$ is omitted in an inter-D2D mode, i.e., $R_i^s \!=\! R^s~\forall i$. In a real system, the average rate can be computed based on the measurements which can be captured by distributions. In this paper, we analyze the average rates for different communication modes by using a stochastic geometry approach~\cite{Lin2013}. Such a stochastic geometry-based performance can serve as the basis for game theory analysis~\cite{Hanawal}.

The coverage probability for a typical user in the $m$-th type mode in the presence of Rayleigh fading is computed as in~\cite{Lin2013} $\mathcal{P}_i^m = \int_0^{\infty} f_m(d)  {e}^{-\!{\sigma^2  \beta_i^m  s_m}} \mathcal{L}_{I_m}(s_m) ~ \mbox{d}d$ where $f_m(d)$ is the probability density function of the link distance, $d$, $\mathcal{L}_{I_m}(s_m)$ is the Laplace transform (LT) of the aggregate interference, $I_m$, $s_m \!=\! \frac{\gamma}{P_ml(d)}$, $P_m$ is a transmit power level, $l(\cdot)$ is the distance-based pathloss model, and $\sigma^2$ is the noise power level calculated over the full cellular band. There are different types of interferers based on D2D sharing approach, i.e., overlay or underlay. For a typical user in the cellular mode, we have $\mathcal{L}_{I_c} \!=\! \mathcal{L}_{I_{cc}}$ in an intra-D2D overlay and $\mathcal{L}_{I_c} \!=\! \mathcal{L}_{I_{cc}} \!+\! \mathcal{L}_{I_{cd}}$ in an intra-D2D underlay where $I_{cc}$ and $I_{cd}$ represent the interference from out-of-cell cellular users and intra-D2D users. For a typical user in the D2D mode, we have $\mathcal{L}_{I_d} \!=\! \mathcal{L}_{I_{dd}}$ in an intra-D2D overlay and $\mathcal{L}_{I_d} \!=\! \mathcal{L}_{I_{dd}}\!+\!\mathcal{L}_{I_{dc}}$ in an intra-D2D underlay where $I_{dd}$ and $I_{dc}$ represent the interference from other intra-D2D users and cellular users. Similarly, in an inter-D2D overlay, we have $\mathcal{L}_{I_s} \!=\! \mathcal{L}_{I_{ss}}$ where $I_{ss}$ represents the interference from other inter-D2D users.

The interference level also depends on the density of interferers computed only after specifying the mode selection scheme. We assume that the locations of \acp{BS}, cellular, intra-D2D and inter-D2D users follow independent Poisson point processes (PPPs) with densities, $\lambda_i^b$, $\lambda_i^c$, $\lambda_i^d$ and $\lambda$, respectively. And the mode selections in intra-D2D and inter-D2D modes just thin the PPPs with $\delta_i$ and $q$. Then, $\mathcal{L}_{I_{cc}}$ and $\mathcal{L}_{I_{cd}}$ can be expressed as in~\cite{Lin2013} $\mathcal{L}_{I_{cc}} \!=\! e^{-2\pi\alpha_i\lambda_{i}^b \int\nolimits_d^\infty \left( \frac{s_c\cdot P_c l(r)}{1 + s_c\cdot P_c l(r) } \right) r dr }$ and $\mathcal{L}_{I_{cd}}  \!=\! e^{-2\pi \delta_i \lambda_i^{d} \int\nolimits_0^\infty \left( \frac{s_c\cdot P_d l(r)}{1 + s_c\cdot P_d l(r) } \right) r dr}$. $\mathcal{L}_{I_{dd}}$ can be obtained by replacing $s_c$ with $s_d$ in $\mathcal{L}_{I_{cd}}$. $\mathcal{L}_{I_{dc}}$ can be obtained with replacing $\delta_i \lambda_i^{d}$ with $\alpha_i\lambda_{i}^b$, $s_c$ with $s_d$ and $P_d$ with $P_c$ in $\mathcal{L}_{I_{cd}}$. $\mathcal{L}_{I_{ss}}$ can be obtained by replacing $\delta_i \lambda_i^{d}$ with $q \lambda$, $s_c$ with $s_s$, and $P_d$ with $P_s$ in $\mathcal{L}_{I_{cd}}$. Note that $\alpha_i$ is the probability a \ac{BS} is active and should take into account not only the densities of cellular users but also the density of D2D users selecting cellular communication mode, i.e., $\!(1\!-\!\delta_i)\lambda_i^d \!+\!(1\!-\!q)\lambda/N$.

\begin{proposition}\label{numerical:concave}
The weighted sum rate utility, $\!U_i \!= \!(1\!-\!w_i^s) \,Q_i^d \!+\! w_i^s \,Q_i^s$, and the weighted \ac{P.F} rate utility, $U_i = (1-w_i^s) \,\log(Q_i^d) + w_i^s \,\log(Q_i^s)$, are concave in $\beta_i$ for $0\leq w_i^s \leq 1$.
\begin{proof}
The proof is in Appendix~\ref{ref:numerical:concave}.
\end{proof}
\end{proposition}

\begin{proposition} \label{numerical:constraint}
The constraint set has an ascending property.
\begin{proof}
The proof is in Appendix~\ref{ref:numerical:constraint}.
\end{proof}
\end{proposition}

Depending on individual intra-D2D mode selection policy, each \ac{OP} may use a two-dimensional strategy domain with a variable internal state, i.e., $\delta_i^{min}\leq {\delta}_i \leq \delta_i^{max}, \forall i$, or a one-dimensional strategy domain with a fixed internal state, ${\delta}_i \rightarrow \delta_i^o$, where $\delta_i^o$ is a fraction of intra-D2D users in D2D mode used for no spectrum sharing. While the utility is concave in $\beta_i$, it is not jointly in $\beta_i$ and $\delta_i$. Due to the ascending property of the strategy space, the solutions of the problem \eqref{eq:P1} will be partly on the boundary of the feasible set, ${\delta}_i \rightarrow \delta_i^{max}$. No matter which intra-D2D mode selection policy is used, i.e., the optimal value of $\delta_i$, ${\delta}_i \rightarrow \delta_i^{max}$, or its fixed value, ${\delta}_i \rightarrow \delta_i^o$, the selected parameter, $\delta_i$ determines the one-dimensional box constraint, and the amount of its coupled spectrum fraction is proposed for the spectrum pool. The solution, $\beta_i$, of the optimization problem \eqref{eq:P1} maximizes the individual utility, but may not be a NE in a non-cooperative spectrum sharing game, $\mathcal{G}$. Next, we study the properties and solutions of the game.

\subsection{\color{black}Existence of NE and Iterative Dynamics}
In $\mathcal{G}$, the NE is an important concept since it represents a steady state where the utilities of all \acp{OP} are maximized. A strategy profile vector $\boldsymbol{\beta}' = [\beta_1',\cdots,\beta_N']\in\mathcal{R}^{N}$ is a NE if for every \ac{OP} $i\in \mathcal{N}$, ${U}_i(\beta_i', \boldsymbol{\beta}_{-i}') \geq {U}_i(\beta_i, \boldsymbol{\beta}_{-i}'), \forall \beta_i \in S_i$ where $\boldsymbol{\beta}_{-i}' \!=\! [{\beta}_{1}'\!,\! \cdots\!,\! {\beta}_{i-1}'\!,\! {\beta}_{i+1}'\!,\!\cdots\!,\!{\beta}_{N}']$. One of the most important questions is whether a NE exists or not.

\begin{proposition}(\cite{Rosen}) \label{prop:NE}
The game $\mathcal{G}$ admits at least one NE, since $S_i$ is a non-empty, compact\footnote{Strategy space is compact if it is closed and bounded.}, and convex set of Euclidean space, and each utility is continuous and concave on $S_i$.
\end{proposition}

Once proven that a NE always exists, the problem of how to reach such an equilibrium arises. In order to reach the NE, an iterative distributed strategy update algorithm can be used. If strategies at $t-1$-th iteration are $\boldsymbol{\beta}^{(t-1)}$, the iterative strategy scheme leads to the strategies for $t$-th iteration, described as $\boldsymbol{\beta}^{(t)} = M(\boldsymbol{\beta}^{(t-1)})$ where $M(\cdot)$ is an iterative response process which is a vector valued mapping function, $M\!\!=\![M_1,\!\cdots,\!M_N]^{T} \subseteq \mathbb{R}^N$. The most common updating orders for $\boldsymbol{\beta}$ based on the mapping $M$ are the parallel update and the sequential update. We prove the convergence for parallel update mapping where all components are updated simultaneously, i.e., $\beta_i^{(t)} = M_i(\boldsymbol{\beta}^{(t-1)}), \forall i$. Due to the lack of space we omit the sequential update case.

In this paper, we consider an iterative algorithm, the \ac{JP} strategy update. The mapping function implemented by the \ac{JP} dynamic is given as 
\begin{equation} \label{eq:jacobi}
M_i(\boldsymbol{\beta}^{(t-1)}) =  (1-\kappa_i^{(t)})\beta_i^{(t-1)} + \kappa_i^{(t)} BR_i(\boldsymbol{\beta}_{-i}^{(t-1)}), \forall i
\end{equation}
where $BR_i(\boldsymbol{\beta}_{-i}) = \arg \max_{\beta_i \in \mathcal{S}_i} U_i(\beta_i,\boldsymbol{\beta}_{-i})$ is the best response to the aggregate proposal from the opponents, $\boldsymbol{\beta}_{-i}^{(t-1)} \!=\! [{\beta}_{1}^{(t-1)}\!,\! \cdots\!,\! {\beta}_{i-1}^{(t-1)}\!,\! {\beta}_{i+1}^{(t-1)}\!,\!\cdots\!,\!{\beta}_{N}^{(t-1)}]$, and $\kappa_i^{(t)}\!\!>\!0$ is called a smoothing parameter\footnote{It is known as speed of adjustment in \cite[p.278]{Moulin1980}. \ac{JP} generally achieves a smoother move than {BR} does in case of non-supermodular games which have a unique NE. The small smoothing parameter plays the role of compensating for the instability of  the {BR} dynamic, see \cite[Sec 4.1.3]{Chen}.} representing the willingness of $i$-th \ac{OP} at $t$-th iteration to maximize its own utility. The \ac{BR} strategy is a special case of the \ac{JP} strategy choosing $\kappa_i^{(t)} \!\!=\!\! 1$. The difference between the two algorithms is in whether or not to have $\kappa_i^{(t)} \!\!=\!\! 1, \forall t$, which enables the \acp{OP} to behave in a myopic manner, strictly or flexibly.

A mapping function satisfying some convergent condition would converge to a NE. This is guaranteed if $M(\cdot)$ is contraction mapping \cite{Moulin1986}, $\rho\left(T^{(t)}\right)\!<\!1$ where $\rho\left(\cdot\right)$ is the spectral radius and $\!T^{(t)} \!\!\!=\!\! \!\left[\frac{\partial M_i(\cdot)}{\partial \beta_j}\right]_{\forall i, j}\;\!\!\!$ are derivatives of $M(\cdot)$. However, when the value of \!$\rho(T^{(t)}\!)$  is used as the convergence criterion, a central entity is needed for collecting the coefficients of $T^{(t)}$ from all \acp{OP}. This undesirable case in $\mathcal{G}$ can be replaced by a sufficient condition.

\begin{lemma}(\cite{Cachon}) \label{lemma:contract2}
If each \ac{OP} satisfies that the sum of the row line of the matrix $T^{(t)}$ is less than one, $\|T^{(t)}\|_{\infty}<1$, the iterative update $\boldsymbol{\beta}^{(t)} = M(\boldsymbol{\beta}^{(t-1)})$ is a contracting iteration.
\end{lemma}

If every \ac{OP} satisfies the condition everywhere individually, and thus the NE obtained as a result of such an iterative play is globally stable, then the iterative response process converges to the unique NE, and thus no matter where the game starts the final outcome will be the same, the global stability of the equilibrium candidate points implies uniqueness. However, this is restrictive, because the contraction mapping may be only satisfied with partial strategy space implying local stability. Thus, the obtained NE might not be unique. In case there are multiple NEs, the selected equilibrium would depend on the initial strategy profile~\cite{Moulin1986}. This might be undesirable because the performance of an \ac{OP} would depend on the initial proposals of other \acp{OP}.

\subsection{{\color{black}Uniqueness of NE}} \label{sec2b}
The non-trivial condition for the uniqueness of a NE in $\mathcal{G}$ can be relaxed by showing that a locally stable NE is unique \cite{Moulin1986}. We check whether a locally stable NE obtained as a result of such an iterative play is unique by the \ac{PH} Index~\cite{Simsek2007,Simsek2005} requiring a certain sign from the Hessian, $H$, but this requirement needs to hold only at a NE, $\boldsymbol{\beta}'$. The obtained sign can be used to define the index for a NE, $Ind(\boldsymbol{\beta}') \!\!=\! \!sign(det(-H(\boldsymbol{\beta}')))$. We present some results on the structure of the NE set, providing conditions for the NEs to be locally unique, finite and globally unique.

Let $\mathcal{K}$ be the set of NE points. We first claim that $\mathcal{K}$, is finite. Consider a NE $\boldsymbol{\beta}'\in \mathcal{K}$ where $\mathcal{K}$ is bounded due to the upper limit of the constraint set and closed due to the continuity of response mapping function, thus compact. The NE is locally unique if it is a strict local maximum of the utility function. If every NE is locally unique, it is locally isolated. Thus, there exists an open neighborhood such that there are no other NEs in the open set. The union of these open sets forms a cover for the set $\mathcal{K}\subset \mathcal{S}$. Since $\mathcal{K}$ is compact, it can be covered by a finite number of sets each containing a NE. Hence, the number of the locally stable NEs, $|\mathcal{K}|$, is finite. Also, due to the fact that the strategy space is non-empty and convex, the sum of the indices for NEs is equal to one, $\sum_{\boldsymbol{\beta}'\in \mathcal{K}} Ind(\boldsymbol{\beta}') =  1$ \cite{Simsek2007}.

A locally stable NE, $\boldsymbol{\beta}'$, is formed by the local maximum of each utility function. A strict local maximum can be ensured by a second order condition \cite{Bertsekas} which requires that i) the NE satisfies the complementary condition, $\nabla U_i(\boldsymbol{\beta}')=0, \forall i$ and ii) the Hessian at complementary NE is negative definite, $det(-H(\boldsymbol{\beta}'))>0$ where $\boldsymbol{\beta}'\in \mathcal{K}$. From the second order conditions, the \ac{PH} index theorem restricts the critical point to be in the interior, $\beta_i^{min} \!<\! \beta_i \!<\!\beta_i^{max}({\delta_i}),\! \forall i$.

In $\mathcal{G}$, it is difficult to rule out the boundary NE where at least one of the \acp{OP} selects the strategy profile on the boundary of the strategy space. The restriction can be resolved by a generalized version of the \ac{PH} index \cite{Simsek2007} which requires complementary and non-degeneracy assumptions. A NE is called non-degenerate if $\nabla U$ is continuously differentiable at $\boldsymbol{\beta}'$ and $H(\boldsymbol{\beta}')|_{\mathcal{\acute{N}}}$ is non-singular where $\mathcal{\acute{N}}= \{i\in \mathcal{N}| \beta_i^{min}<{\beta_i'} < \beta_i^{max}({\delta_i})\}$ denotes the set of \acp{OP} selecting interior profile, and $H|_{\mathcal{\acute{N}}}$ denotes the principal sub-matrix of $H$ corresponding to the indices in ${\mathcal{\acute{N}}}$. Thus, the non-degeneracy assumption boils down to the complementary condition for the sub-matrices of $H$.

In $\mathcal{G}$, it is still non-trivial to evaluate the sign of the sub-matrix of the Hessian at a non-degenerate point. When an \ac{OP} chooses the lower limit, $\beta_i'=\beta_i^{min}$, or upper limit, $\beta_i'=\beta_i^{max}(\delta_i)$, of his own strategy, other \acp{OP} are not able to notice the boundary profile. Thus, it is difficult to obtain the sub-matrix structure, $H(\boldsymbol{\beta'})|_{\mathcal{\acute{N}}}$, and check the non-singularity of the sub-matrix. In \cite{Simsek2005}, a stronger non-degeneracy condition is introduced to replace the complementarity condition with $P$-matrix property where the determinants of the arbitrary principal sub-matrices are positive.

When $-H(\boldsymbol{\beta}')$ is a $P$-matrix and $\nabla U$ is continuously differentiable at $\boldsymbol{\beta}'\in \mathcal{K}$, we say that $\boldsymbol{\beta}'$ is a $P$-critical point. The critical point is not necessarily complementary, if the $P$-matrix property holds. Every $P$-critical point is non-degenerate. This fact allows accounting critical points on the boundary for their contribution to the index sum.

\begin{lemma} \label{lemma_rowdiag}
A NE, $\boldsymbol{\beta}'\in \mathcal{K}$, has a positive one index, if $-H(\boldsymbol{\beta}')$ is  diagonally dominant with positive diagonal elements.
\begin{proof}
The proof is in Appendix \ref{ref:lemma_rowdiag}.
\end{proof}
\end{lemma}

\begin{remark} \label{remark1}
The NE is unique if and only if every NE, $\boldsymbol{\beta'}\in \mathcal{K}$ has a positive one index.
\begin{proof}
If multiple NEs, each of which fulfilling {\color{black}Lemma~\ref{lemma_rowdiag}}, exist, the sum of the indices is equal to the number of the NEs, $\!\!\sum_{\boldsymbol{\beta}'\!\in \mathcal{K}}\!\! Ind(\boldsymbol{\beta}') \!=\!\!  |\mathcal{K}|$. This contradicts $\sum_{\boldsymbol{\beta}'\!\in \mathcal{K}} \!Ind(\boldsymbol{\beta}') \!=\! \!1$.
\end{proof}
\end{remark}

\subsection{Connection between Uniqueness and Local stability of NE}
The following proposition shows that the Hessian matrix can be expressed in terms of the jacobian matrix, $T$ from the implicit function theorem \cite{Cachon}. This would allow us to verify the condition for the row diagonal dominant matrix property ({\color{black}Lemma \ref{lemma_rowdiag}}) with the sufficient condition for the local contraction ({\color{black}Lemma \ref{lemma:contract2}}).
\begin{proposition} \label{prop_Hfun}
Hessian matrix, $H(\boldsymbol{\beta}')$ can be expressed in terms of the jacobian matrix, $T$, as $H(\boldsymbol{\beta}') = D(I-T)$ where $D$ is a $N\times N$ diagonal matrix with $[\frac{\partial^2 U_i}{\partial {\beta_i}^2}]_{i}$, $I$ is a $N\times N$ identity matrix and $T$ is a $N\times N$ matrix $[{\partial M_i}/{\partial {\beta_j}}]_{i,j}$. 
\begin{proof}
The proof is in Appendix \ref{ref:prop_Hfun}.
\end{proof}
\end{proposition}

The following lemma shows that the row diagonally dominant property of the matrix $-H(\boldsymbol{\beta}')$ is equivalent to the condition that the matrix norm induced by the infinity norm is less than one at the NE point $\boldsymbol{\beta}'$, $\|T\|_{\infty} <1$.

\begin{lemma} \label{lemma_index5}
A NE, $\boldsymbol{\beta}'$, has a positive one index, $Ind(\boldsymbol{\beta}')= 1$, if $U_i$ is concave in $\beta_i$ and $\|T\|_{\infty}<1$. 
\begin{proof}
The proof is in Appendix~\ref{ref:lemma_index5}.
\end{proof}
\end{lemma}

\begin{remark} \label{remark11}
If every NE is locally stable (Lemma \ref{lemma_index5}), $\mathcal{K}$ has an exact point, its uniqueness (Remark \ref{remark1}).
\end{remark}

\begin{remark} \label{remark:2}
Local stability of NE is equivalent to its uniqueness ({\color{black}Remark \ref{remark11}}), but the opposite is not necessarily true.
\end{remark}

Recall that to prove that the sub-matrix $\!-H(\boldsymbol{\beta}')$ is a $P$-matrix, the determinants of arbitrary principal sub-matrices $-H(\boldsymbol{\beta}')|_{{\mathcal{\bar{N}}}}$ for any ${\mathcal{\bar{N}}}\!\!\subseteq\!\! {\mathcal{{N}}}$ should be positive, $det(-H(\boldsymbol{\beta}')|_{\mathcal{\bar{N}}})>0$. The determinant of a sub-matrix $-H(\boldsymbol{\beta}')|_{\mathcal{\bar{N}}}$ can be expressed as, according to {\color{black}Prop. \ref{prop_Hfun}}, $det(-H(\boldsymbol{\beta}')|_{\mathcal{\bar{N}}})\!=\!det(-\bar{D} (\bar{I}-\bar{T})) \!=\! det(-\bar{D}) det(\bar{I}-\bar{T})$ where a matrix $\bar{A}$ denotes the principal sub-matrix of a matrix $A$ corresponding to the indices in ${\mathcal{\bar{N}}}$. The first matrix, $-\bar{D}$, is a ${\mathcal{\bar{N}}}\times {\mathcal{\bar{N}}}$ diagonal matrix whose elements are positive due to the concavity of the utility function, $\frac{\partial^2 U_i}{\partial \beta_i^2} <0, \forall i$. To estimate the determinant of the second matrix, $\bar{I}-\bar{T}$, we first let $\bar{\xi}_i, \forall i$ be the eigenvalues of the jacobian matrix $\bar{T}$. The eigenvalues and the determinant of the second matrix, $\bar{I}- \bar{T}$, are $1-\bar{\xi}_i$ and $\prod_i (1-\bar{\xi}_i), \forall i\in {\mathcal{\bar{N}}}$. 
Thus, we have a positive determinant, $det(\bar{I}-\bar{T}) \!>\!0$, if $\bar{\xi}_{i} <1,\! \forall i\in {\mathcal{\bar{N}}}$.

\begin{remark} \label{remark:3}
Local instability of NE could occur, even though there is a unique NE.
\begin{proof}
If the minimum and maximum eigenvalues of $\bar{T}$ are less than a negative one and a positive one, respectively, $\min{(\bar{\xi}_i)}<-1$ and $\max{(\bar{\xi}_i)}<1, ~\forall i$, for $\frac{\partial^2 U_i}{\partial \beta_i\beta_j}<0, \forall{\mathcal{\bar{N}}}\subseteq {\mathcal{{N}}}$, then according to Remark \ref{remark:2}, the game $\mathcal{G}$ has a unique NE. However, due to a negative dominant eigenvalue less than $-1$, it has $\rho(\bar{T})>1$. This implies that there is a unique but unstable NE, thus divergence of $M(\cdot)$.
\end{proof}
\end{remark}

Due to the fact that the spectral radius of a matrix is bounded by its matrix norm and the infinity norm property, we have $ \rho(\bar{T})\!\!\leq\!\!\|\bar{T}\|_{\infty}\!\!\leq\!\!\|{T}\|_{\infty}$. The {\color{black}Remarks \ref{remark:2} and \ref{remark:3}} motivate us to identify a proper contraction mapping resulting in $\|T\|_{\infty}\!\!<\!\!1$ for convergence to a unique NE.

\section{Convergence of Jacobi-play dynamic}
\label{sec:algorithms}
In this section, we provide necessary and sufficient conditions for the convergence of \ac{JP} dynamic to a unique NE, and propose a distributed algorithm where each \ac{OP} checks the conditions independently and makes an offer about the amount of spectrum committed to the spectrum pool. Then the offers are exchanged and so forth till consensus is reached. While making the offer, each \ac{OP} considers only its individual reward based on the opponents' proposals, and does not reveal any operator-specific information to the opponents.

\subsection{Sufficient condition for Convergence}
To analyze the convergence of the \ac{JP} dynamic in (\ref{eq:jacobi}), we consider the jacobian matrix $T^{(t)} \!=\! {J}^{(t)}$ of the self-mapping function in (\ref{eq:jacobi}) and the elements of the jacobian matrix ${J}^{(t)}$ are $ J_{ij}^{(t)} \!=\! \kappa_i^{(t)} J_{ij}^{BR(t)}~\mbox{for}~ i\!\neq\! j$ and $ J_{ij}^{(t)} \!=\! 1\!-\!\kappa_i^{(t)}~\mbox{for}~ i \!=\! j$. When $\kappa_i^{(t)} \!=\! 1, \forall t$, the jacobian matrix ${T}^{(t)}$ corresponds to the \ac{BR} dynamic, ${J}^{(t)}  \!=\! {J}^{BR(t)}$ with element $J_{ij}^{BR(t)}$ denoting the slope of the \ac{BR} of $i$-th \ac{OP} to the strategy profile of the $j$-th \ac{OP}: $J_{ij}^{BR(t)} \!\!=\!\! \frac{\partial BR_i(\boldsymbol{\beta}_{-i}^{(t-1)})}{\partial \beta_j} ~\mbox{for}~ i\neq j$ and $J_{ij}^{BR(t)} \!=\! 0 ~\mbox{for}~ i = j$. When $\rho({J}^{BR(t)})<1$, then the \ac{BR} converges to the unique NE. A sufficient condition for the \ac{BR} update function to exhibit a contraction mapping is to show that the maximum absolute row sum matrix norms is less than one, $\sum_{j\neq i} |J_{ij}^{BR (t)}| \!< \!1, \forall i$ ({Lemma \ref{lemma:contract2}}). When $\rho({J}^{BR(t)})\!>\!1$, then the \ac{BR} would diverge. However, if the maximum eigenvalue for ${J}^{BR(t)}$ is less than a positive one, there is a unique NE as discussed after {Remark \ref{remark:2}}. When $\kappa_i^{(t)} \!\neq\! 1$, ${J}^{(t)}$ is different from ${J}^{BR}$. If $\rho({J}^{(t)})\!<\!1$ for $t\geq 0$, the \ac{JP} converges to the unique NE. Otherwise, it would diverge, but the stability can be ensured by selecting a proper $\kappa_i^{(t)},\forall i$, at each $t$ as we will show later. The presence of the diagonal terms will make the stability conditions in general different.

\begin{remark} \label{remark:br:dominant}
If the matrix ${J}^{BR(t)}$ has the dominant eigenvalue less than $-1$ and the other eigenvalues less than $1$, then according to {\color{black}Remark \ref{remark:3}}, there is a unique but unstable NE for the \ac{BR} dynamic. The instability of the \ac{BR} dynamic can be compensated by $\kappa_i^{(t)}<1,\forall i$.
\end{remark}

\begin{proposition} \label{prop:JC}
When the \ac{BR} dynamic converges to the unique NE, $\sum_{j\neq i} |J_{ij}^{BR{(t)}}|\!<\!1, \forall i$, the \ac{JP} also converges with $0< \kappa_i(t)< {2}/({1+\sum_{j\neq i} |J_{ij}^{BR(t)}|})$.
\begin{proof}
The proof is in Appendix~\ref{ref:prop:JC}.
\end{proof}
\end{proposition}

We have shown that the \ac{JP} dynamic converges, as long as a proper $\kappa_i^{(t)}$ is selected. However, the sufficient condition for convergence does not hold, if $\sum_{j\neq i} |J_{ij}^{BR(t)}|\!\geq\!1$. According to \cite{Kahan}, a necessary condition for the \ac{JP} scheme to converge is $|1-\kappa_i^{(t)}|\!<\!1$, equivalently, $0\!<\! \kappa_i^{(t)}\!<\! 2$, which holds for $\sum_{j\neq i} |J_{ij}^{BR(t)}|\!<\!1$, but does not hold for $\sum_{j\neq i} |J_{ij}^{BR(t)}|\!\geq\!1$. Another condition is derived for $\rho({J}^{BR(t)})\!>\!1$ in \cite[Theorem 2.2]{Hadji}. However, the condition is quite restrictive, since all of the eigenvalues of the jacobian matrix ${J}^{}$ should be with non-positive real parts and the smoothing parameters should be identical, $\!\kappa_i^{(t)} \!=\! \kappa^{(t)}\!, \forall i$. For stability analysis in $\mathcal{G}$, the parameter $\kappa_i^{(t)}$ is not known to other \acp{OP} and it is non-trivial to find the exact eigenvalues of ${J}^{(t)}$.

The eigenvalues of a matrix can be estimated by using Gerschgorin's circle theorem~\cite{Ortega} which provides bounds on the eigenvalues. Every eigenvalue of the matrix $J^{(t)}$ lies within the union of discs, ${\xi} \subseteq \bigcup_{i=1} \rho_i$, where ${\xi}$ denotes the eigenvalues of ${J}^{(t)}$ and $\rho_i$ is a Gerschgorin's circle with center $J_{ii}^{(t)}$ and radius $\sum_{j\neq i} |J_{ij}^{(t)}|$, $\!\rho_i \!\!=\!\! \{ z\!:\! \!|z-J_{ii}^{(t)}|\!<\! \sum_{j\neq i} |J_{ij}^{(t)}|\}$. If every disk is inside the unit circle, every eigenvalue lies in the union of the disk and thus $\!\rho(J^{(t)})\!<\!1$. Since the diagonal element in each row of the matrix $J^{(t)}\!\!$ is real, the center of each circle lies on the x-axis. The eigenvalues are bounded such that $ - \sum_{j\neq i} |J_{ij}^{(t)}| \!<\! {\xi} \!- \!J_{ii}^{(t)} \!<\! \sum_{j\neq i} |J_{ij}^{(t)}|, \forall i$ and thus $ - \sum_{j\neq i} |J_{ij}^{(t)}|+ J_{ii}^{(t)} < {\xi} < \sum_{j\neq i} |J_{ij}^{(t)}| + J_{ii}^{(t)}, \forall i$. If the lower and upper limits of the circle region are larger than a negative one and less than a positive one, respectively,  $ -1< -\kappa_i^{(t)}\sum_{j\neq i} |J_{ij}^{BR(t)}|+ (1-\kappa_i^{(t)})$ and $\kappa_i^{(t)}\sum_{j\neq i} |J_{ij}^{BR(t)}|+ (1-\kappa_i^{(t)}) <1$, $\forall i$, the modulus of every eigenvalue of the matrix $J^{(t)}$ is strictly less than one.

\begin{lemma}\label{lemma:jc:dominant}
If the maximum eigenvalue of ${J}^{(t)}$ is less than one, the small value of $\kappa_i^{(t)}$ in the Jacobi update can compensate for the instability of the \ac{BR} dynamics.
\begin{proof}
The proof is in Appendix~\ref{ref:lemma:jc:dominant}.
\end{proof}
\end{lemma}

\subsection{\color{black}Necessary and sufficient condition for Convergence}
Next, we find the necessary and sufficient condition for convergence to a unique NE in $\mathcal{G}$. For this, we show that all eigenvalues in absolute value are less than one. This condition is satisfied if the dominant eigenvalue is negative, larger than a negative one. According to \cite[Theorem 2]{James}, the jacobian matrix, ${J}^{(t)}$, has exactly one negative eigenvalue with every other eigenvalue having no larger modulus, if all principal minors are negative. However, finding the condition on all negative principal minors of ${J}^{(t)}$ is also difficult in $\mathcal{G}$.

This non-trivial work can be resolved by recalling that in spectrum pooling the utility, $U_i$, is impacted by additive combinations of the spectrum fractions other \acp{OP} contribute to the pool, $\beta \!=\! \sum_{i} \beta_i$. Due to the linearity, the off-diagonal elements at row $i$ in the jacobian matrix, $J_{ij},j\neq i$, are identical, equivalently the slopes of each \ac{OP}'s response function with respect to each opponent are identical.

\begin{lemma} \label{prop:pooling}
In $\mathcal{G}$, let an $N \times N$ matrix ${J}^{} \!\!=\!\! [{J}_{ij}]_{i,j \in N}$ have $J_{ij}\!\!=\!\!J_{i}$, $\forall j\neq i$, and let all $J_{i}^{}$ have the same signs. When $J_{i}^{}\!\!<\!\! 0, \forall i$, the necessary and sufficient conditions for all eigenvalues of ${J}$ to have values less than unity in absolute form are $ \sum_i \frac{-J_{i}^{}}{(1+J_{ii} - J_i^{})} \!<\! 1$ and $|J_{ii}-J_i^{}|\!<\! 1$.
\begin{proof}
The proof is in Appendix~\ref{ref:prop:pooling}.
\end{proof}
\end{lemma}

\begin{proposition} \label{lemma:pooling2}
In $\mathcal{G}$, there is a unique NE if $-1<J_{ij}^{BR}<0$, $j\neq i, \forall i,j$.
\begin{proof}
The proof is in Appendix~\ref{ref:lemma:pooling2}.
\end{proof}
\end{proposition}

Based on {\color{black}Prop. \ref{lemma:pooling2}}, every \ac{OP} can inform its peers whether $-1\!<\!J_{ij}\!<\!0, \forall j$. If all indications are positive, the \acp{OP} identify in a distributed manner that the NE is unique.

\begin{proposition} \label{lemma:pooling3}
In $\mathcal{G}$, \ac{BR} converges to the unique NE, if $-\frac{1}{N-1}<J_{ij}^{BR}<0$ $j\neq i, \forall i,j$.
\begin{proof}
The proof is in Appendix~\ref{ref:lemma:pooling3}.
\end{proof}
\end{proposition}

\begin{proposition} \label{lemma:pooling4}
In $\mathcal{G}$, \ac{JP} converges to the unique NE, if $0<\kappa_i^{(t)}<\kappa_{i,max}^{(t)}$ where $\kappa_{i,max}^{(t)} =\frac{2}{1+ (N-1) |J_{ij}^{BR}|}$ and $-1<J_{ij}^{BR}<0$ for $t \geq 0$, $j\neq i$.
\begin{proof}
The proof is in Appendix~\ref{ref:lemma:pooling4}.
\end{proof}
\end{proposition}

\begin{figure}[t]
\centering\vspace{-20pt}
\includegraphics [width=.975\linewidth]{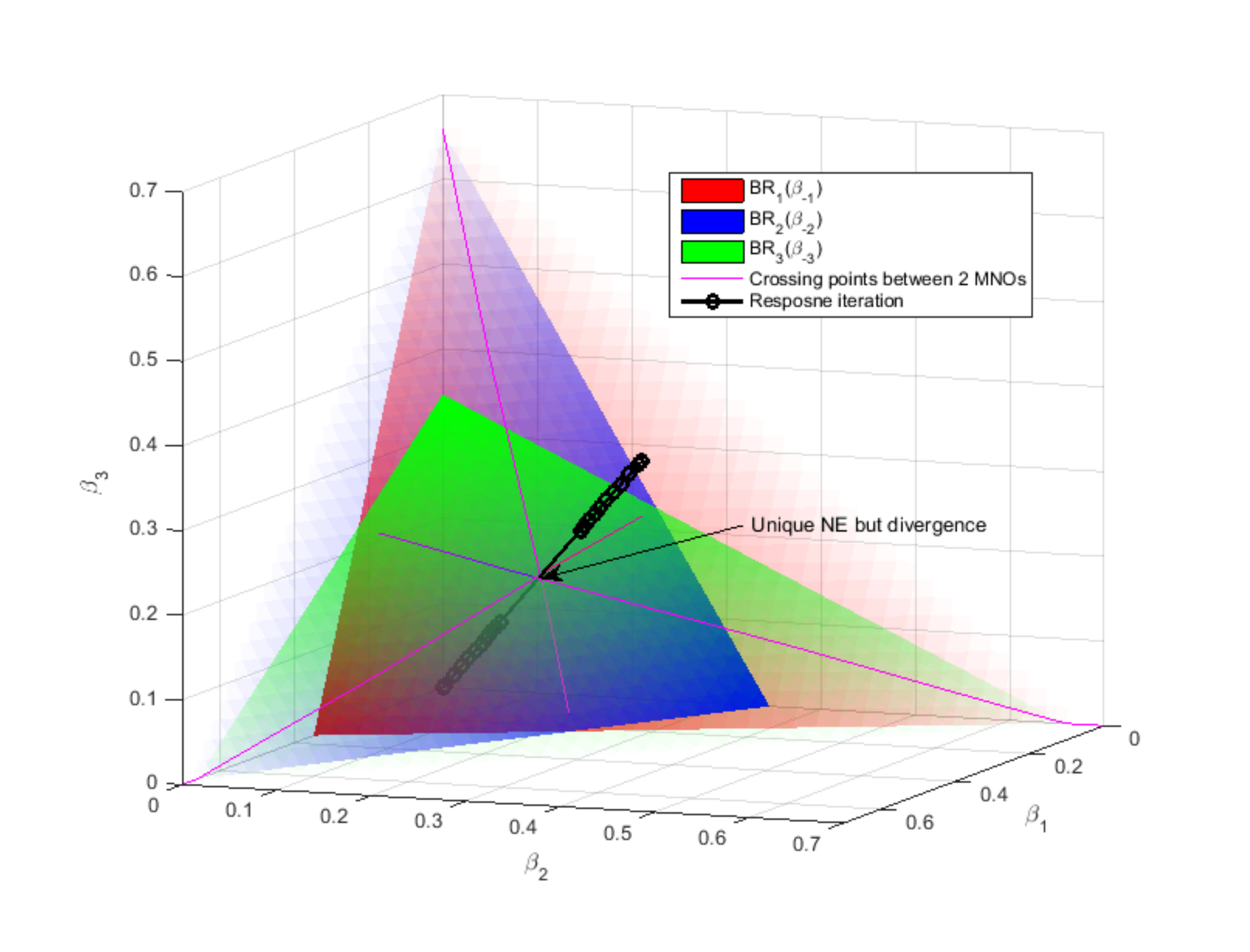}\vspace{-15pt}
 \caption{Illustration of NE divergence of \ac{BR} for $N =3$. Shaded area shows \ac{BR} strategy profiles responding to opponents' aggregate proposal: There exists a unique NE at crossing point but the \ac{BR} diverges. MNO denotes mobile network operator.}
\label{Numerical:fig:global}
\vspace{-1em}
\end{figure}

Intuitively, the use of $\kappa_i^{(t)}$ within the range $(0, \kappa_{i, max}^{(t)})$ is helpful in enabling convergence since it prevents the overreacting response to the proposals from the opponents. {\color{black}Fig. \ref{Numerical:fig:global}} shows an example indicating that even if there is a unique fixed point, the \ac{BR} dynamic does not converge to the desired point, but the \ac{JP} dynamic can converge with a suitable smoothing parameter $\kappa_i^{(t)}$. Since each \ac{OP} behaves in a myopic manner, it would select the maximum available parameter. Since $\kappa_i^{(t)}$ should be strictly less than $\kappa_{i,max}^{(t)}$ (Prop. \ref{lemma:pooling4}), selecting $\kappa_i^{(t)}\!=\! \kappa_{i,max}^{(t)}$ results in divergence of the \ac{JP} dynamic. Therefore, each \ac{OP} needs to select a lower value than $\kappa_{i,max}^{(t)}$, $\bar{\kappa}_{i}^{(t)}$ which can be achieved by using an upper bound on $|J_{ij}^{BR}|$. From Eq. (\ref{eq:implicit_func}) and ${J}^{BR}$, the upper bound on $|J_{ij}^{BR}|$ can be obtained from an upper bound on $\left|\frac{\partial^2 U_i(\cdot)}{\partial \beta_i\partial \beta_j}\right|$ which depends on the utility function, $U_i$, based on the aggregate spectrum fraction from the opponents, $\boldsymbol{\beta}_{-i}^{(t)}$.

\begin{algorithm}[h]
\caption{Jacobi-play strategy update}  \label{algorithm1}
\begin{algorithmic}[1]  \small
\State $t \leftarrow 0$, Initialize ${\beta}_i^{(t)} \in S_i, \forall i \in\mathcal{N}$
\Repeat 
\For{$i\in \mathcal{I}$} 
\State $\beta_i^{(t+1)} = BR_i(\boldsymbol{\beta}_{-i}^{(t)}; \delta_i^{max}, q)$
\If {$\beta_i^{(t+1)} \notin$ contraction region}
\State $\kappa_i^{(t)}\leftarrow \bar{\kappa}_i^{(t)}$
\Else 
\State $\kappa_i^{(t)}\leftarrow 1$
\EndIf
\State $\beta_i^{(t+1)}\!=\!(\!1\!-\!\kappa_i^{(t)}\!)\beta_i^{(t)} \!+\! \kappa_i^{(t)}
\beta_i^{(t+1)}$
\State $t \leftarrow t + 1$
\EndFor
\Until{Convergence}
\end{algorithmic}
\end{algorithm}

%

\subsection{Proposed distributed algorithm}
In $\mathcal{G}$, neither the utility functions nor the precise outcome levels of the utility functions of other opponents need be known to each \ac{OP}. It is only necessary that each \ac{OP} knows the behavior of the actual proposals the other \acp{OP} contribute to the spectrum pool. Given the \ac{BR} update, if the contraction condition is satisfied\footnote{When $\sum_{j\neq i} |J_{ij}^{BR}|\!<\!1$, $\kappa_{i,max}^{(t)} > 1$. Thus, the consequence of the myopic manner might result in slow convergence rate.}, the \ac{JP} uses $\kappa_i^{(t)}=1$, acting like the \ac{BR} update. If the condition is not satisfied, $\kappa_{i,max}^{(t)}$ becomes less than 1, and $\bar{\kappa}_i^{(t)}$ is chosen for $\kappa_i^{(t)}$\footnote{Convergence speed depends on how close dominant eigenvalue is to 0 \cite{Ortega}, i.e., $\!{\kappa}_i^{(t)} \!\!=\!{1}/({\sum_{j\neq i}\!|J_{ij}^{BR(t)}\!|}\!+\!1), \forall i$, (Lemma \ref{lemma:jc:dominant}).}. To sum up, we set $\kappa_i^{(t)}=1$ if $\frac{-1}{N-1}\!<\!J_{ij}^{BR}\!<\!0$  ({\color{black}Prop. \ref{lemma:pooling3}}) and $\kappa_i^{(t)}=\bar{\kappa}_{i}^{(t)}$ if $-1\!<\!J_{ij}^{BR}\!\leq\! {-1}/({N-1})$ ({\color{black}Prop. \ref{lemma:pooling4}}). For a possible algorithm implementation, see Algorithm \ref{algorithm1}. Note that the \ac{JP} strategy takes place only if there is a unique NE in $\mathcal{G}$. The uniqueness of NE is ensured if every \ac{OP} who has a concave utility on a box-constrained region fulfills the sufficient condition in Prop. \ref{lemma:pooling2}. The sufficient condition can be verified distributively among the \acp{OP} who are willing to participate in $\mathcal{G}$. An \ac{OP} who does not fulfill the sufficient condition would not participate in $\mathcal{G}$, since its participation could cause the existence of multiple NEs which is undesirable in \ac{CoPSS}. We assume that the iteration converges much faster than any change detected in the channel.

After the iterative distributed algorithms result in converged NE, it is natural to assume that the agreement will break if the utility of an \ac{OP} is lower than the utility corresponding to no spectrum sharing. In \cite{Zhong2014} and \cite{Zhong2015}, distributed algorithms in non-cooperative resource allocation games were proposed to obtain a NE. However, the strategy update step size which similarly acts as $\kappa_i^{(t)}$ in this paper is a predetermined constant~\cite{Zhong2014} or is determined by central entity~\cite{Zhong2015}. Such algorithms are not applicable to \ac{CoPSS} scenario, since they do not ensure the stability of NE in a distributed manner.

The converged NE is a point where one is likely to end up operating after the participating \acp{OP} compete with one another. However, even though the NE is unique, in general, it is not an pareto-optimal, or even a desirable solution from a social point of view. Thus, it is worth evaluating the efficiency of the converged NE, enabled by a comparison with the solution yielding a social welfare maximization, i.e., $\boldsymbol{\beta}^*= \arg\max_{\boldsymbol{\beta} \in \mathcal{S}}\sum_{i\in\mathcal{N}}U_i$. Denote $\psi= \frac{\sum_{i\in \mathcal{N}} U_i(\boldsymbol{\beta}')}{\sum_{i\in \mathcal{N}} U_i(\boldsymbol{\beta}^*)}$ by the ratio of the sum of the utilities at the converged NE, $\boldsymbol{\beta}'$ to the one at the pareto-optimal point, $\boldsymbol{\beta}^*$. The value of $\psi$ indicates how the efficiency of the NE solution degrades due to the selfish behavior of \acp{OP} in $\mathcal{G}$, i.e., $\psi$ closer to $1$ is socially better. In this paper, the socially optimal solution is obtained for the sake of comparison and study of NE efficiency.

\acp{OP} may disagree to operate at the social optimal solution, if each utility at $\boldsymbol{\beta}^*$ is lower than the one at $\boldsymbol{\beta}'$, i.e., they want to cooperate but nevertheless act with self-interest. In this case, a cooperative solution based on the Nash product~\cite{Peters} can be computed, i.e., $\arg\max_{\boldsymbol{\beta} \in \mathcal{S}} \prod_{i\in\mathcal{N}} (U_i - U_{i,d})$ where $U_{i,d}$ is the disagreement. The converged NE can play a treat point in such a cooperative game, since it represents the outcome in the event the \acp{OP} would realize their threat not to cooperate. With this disagreement, we may restrict the search space for cooperative solutions to the sub-region consisting of all points of $\mathcal{S}$, making the search space smaller than the case when $U_{i,d} \!=\! 0$ or $U_{i,d} \!=\! U_{i,o}$ where $U_{i,o}$ is the utility without \ac{CoPSS}, i.e., \acp{OP} who do not meet the sufficient condition in Prop.6 may agree to operate cooperatively with $U_{i,d} \!=\! 0$ or $U_{i,d} \!=\! U_{i,o}$.

\section{Numerical illustration}\label{sec:Numerical}
\subsection{Parameter settings} 
Each \ac{OP} is assumed to have BSs with a density equal to inter-site distance of $500$ m \cite{{3GPP_EUTRA}} and cellular users with 5 times the BS density. The distributions of \acp{BS} and cellular users are independent. A cellular user is associated with the nearest home-operator \ac{BS}. The intra-D2D and inter-D2D users are randomly distributed with densities, $\lambda_i^d$ and $\lambda = \sum_i \lambda_i$. In the numerical illustration, we use $q \! = \! \delta_i^{max} \!\!=\!\! 1$ and thus the densities are assumed to be the ones after a mode selection. We take a 3GPP propagation environment~\cite{3GPP_EUTRA} into account with a distance-based pathloss function $l(r)$ in dB: $37.6 \log_{10}(r) \!+\! 15.3$ for the cellular mode and, $40.0\log_{10}(r) \!+\! 28$ for the D2D mode, where $r$ is the distance in meters. The D2D link distance is fixed to $d\!=\!10$ m. We use fixed transmit power levels equal to  $10$ dBm for the D2D mode and $23$ dBm for the cellular mode~\cite{3GPP2014}. The target rates for cellular users and intra-D2D users in D2D mode are $\tau_i^c\!=\!0.1$ and $\tau_i^d\!=1.0, \forall i$. Each \ac{OP} contributes the positive spectrum fraction, $\beta_i^{min}=0.01$.

\begin{figure*}[t]
\centering
\begin{tabular}{c c c} \hspace{-20.2pt}
 \subfigure[\label{Numerical:fig:utility}]{\includegraphics[width=0.335\textwidth]{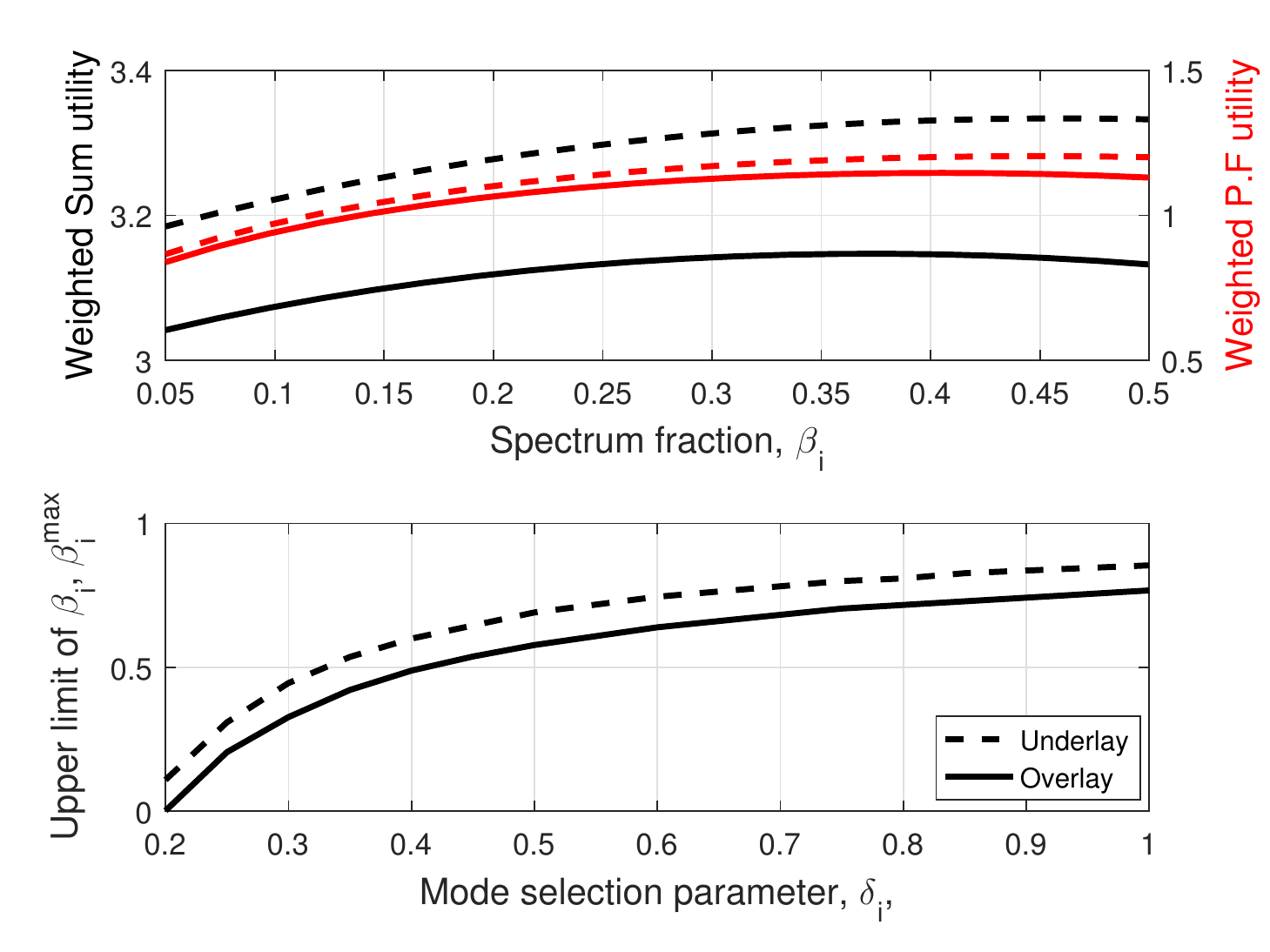}}& \hspace{-15pt}
 \subfigure[\label{Numerical:fig:JC}]{\includegraphics[width=0.335\textwidth]{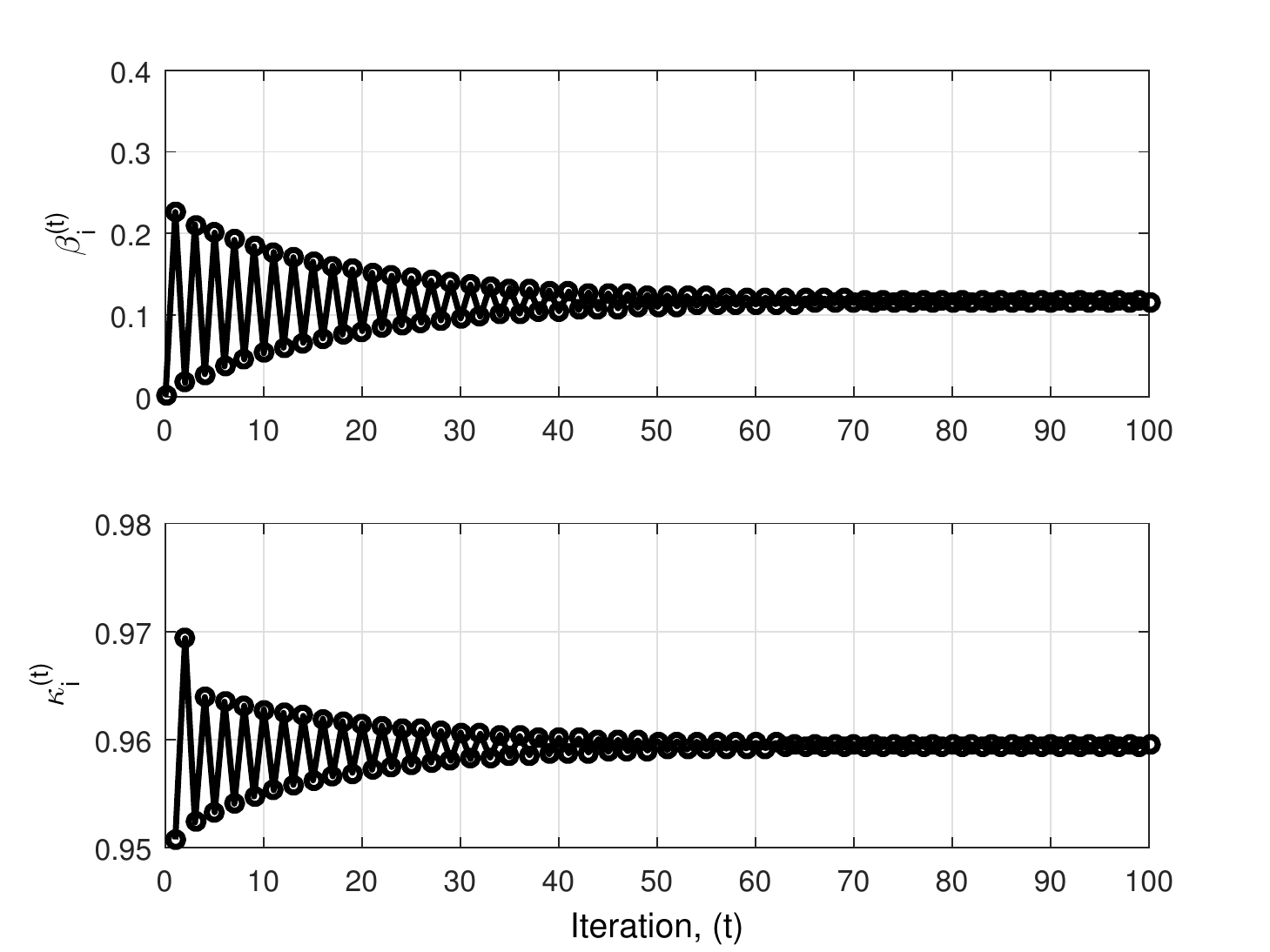}}& \hspace{-15pt}
 \subfigure[\label{Numerical:fig:BR}]{\includegraphics[width=0.335\textwidth]{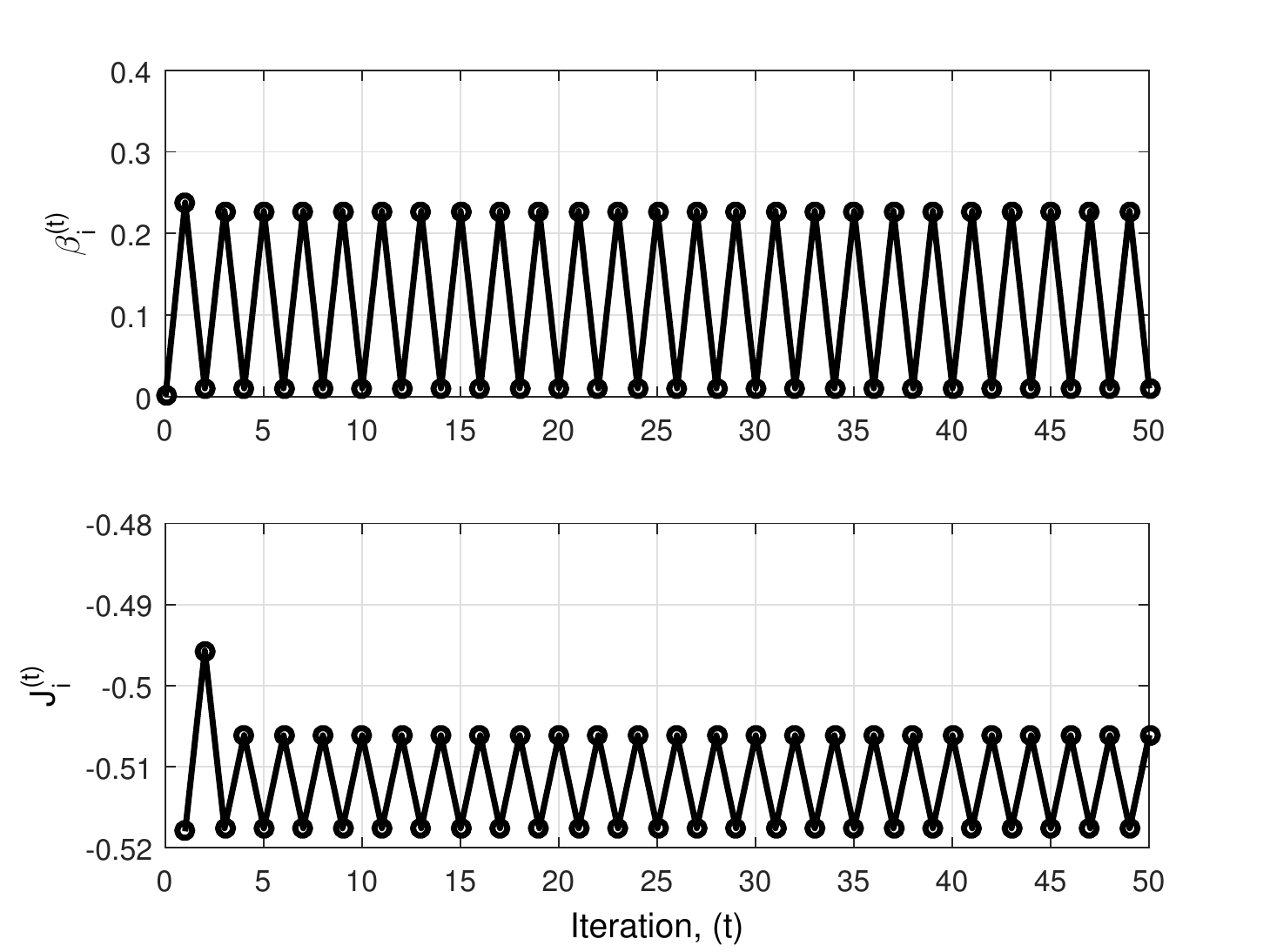}} 
\end{tabular} 
 \caption{(a) {Concavity and monotonicity}, (b) Convergence of \ac{JP}, and (c) Divergence of \ac{BR} to a unique NE, when $N =3$.}
\end{figure*}

\subsection{Convergence}
Fig. \ref{Numerical:fig:utility} shows the concavity of the utilities with respect to $\beta_i$, proven in Prop. \ref{numerical:concave}, and the monotonic property of the constraint set with respect to $\delta_i$, proven in Prop. \ref{numerical:constraint}. As the optimal intra-mode selection, we have $\delta_i \!\rightarrow\! \delta_i^{max}\!\!$ at each iteration, due to Prop \ref{numerical:constraint}. Thus, from Props. \ref{numerical:concave}, \ref{numerical:constraint}, every \ac{OP} has a concave utility on the box-constrained region, $\beta_i^{min}\leq \beta_i \leq \beta_i^{max}(\delta_i^{max})$. From Prop. \ref{lemma:pooling2}, the uniqueness of NE is guaranteed if every \ac{OP} meets $-1 <J_{ij}^{BR} <0, \forall i$.

\begin{proposition}\label{numerical:uniqueness2}
The uniqueness of NE is guaranteed, if every \ac{OP} has a concave utility w.r.t $\beta_i$.
\begin{proof}
The proof is in Appendix \ref{ref:numerical:uniqueness2}.
\end{proof}
\end{proposition}

\begin{proposition}\label{numerical:dominance}(\cite{Cho2015})
The BR converges to the unique NE for $\!N\!=\!2$, if the DSC, $\!\left|\frac{\partial^2 U_i}{\partial\beta_i^2}\right| \!\!>\!\! \left|\frac{\partial^2 U_i}{\partial\beta_i \partial\beta_j}\right|,\!\!\forall i, j \in\{1,2\}$, is satisfied.
\end{proposition}

\begin{proposition} \label{lemma:JC}
The \ac{JP} converges to the unique NE, if $\kappa_i^{(t)}$ is set to be less than or equal to $\bar{\kappa}_{i}^{(t)} = {2}/({1+(N-1) |\bar{J}_{ij}^{BR}|})$ where $|\bar{J}_{ij}^{BR}|$ is an upper bound to $|J_{ij}^{BR}|^{}$. 
\begin{proof}
The proof is in Appendix \ref{ref:lemma:JC}.
\end{proof}
\end{proposition}

Fig.~\ref{Numerical:fig:JC} shows an example on the convergence of \ac{JP} dynamic to a unique NE for symmetric \acp{OP}. For robust convergence, each \ac{OP} independently selects a proper relaxed $\bar{\kappa}_{i}^{(t)}$, compensating for the instability of the \ac{BR} dynamic. At $t \!=\! 1$, each \ac{OP} finds the strategy profile, $\beta_i^{(1)} \!=\! 0.24$, which is not in the contraction region. As shown in Algorithm $1$, each \ac{OP} uses the bounded parameter $\bar{\kappa}_{i}^{(t)} \!=\! 0.95$ as ${\kappa}_{i}^{(t)}$, and updates the strategy profile resulting in $\beta_i^{(1)} = 0.22$ which is in the contraction region. Then, the strategy profiles will be set in a similar way for the next iteration and so forth. In the end, the \ac{JP} dynamic converges to the unique NE, $\beta_i' = 0.12$. Fig.~\ref{Numerical:fig:BR} shows a divergence of the \ac{BR} for the same parameter setting as in Fig.~\ref{Numerical:fig:JC}. The \ac{BR} is not a contraction due to $J_{i}^{(t)} \!=\! J_{ij}^{BR(t)} \!<\!-0.5$ for $N\!=\!3$ \acp{OP} and $t\geq 0$. It should be nevertheless noted that $J_{i}^{(t)}$ are between $-1$ and $0$, $\forall t\geq 0$, ensuring that there exist a unique NE ({\color{black}Props.~\ref{lemma:pooling2} and~\ref{lemma:pooling3}}).

\begin{figure*}[t]
\centering
\begin{tabular}{c c c} \hspace{-13.2pt}
\subfigure[\label{Numerical:fig:asym:a}]
{\includegraphics[width=0.335\textwidth]{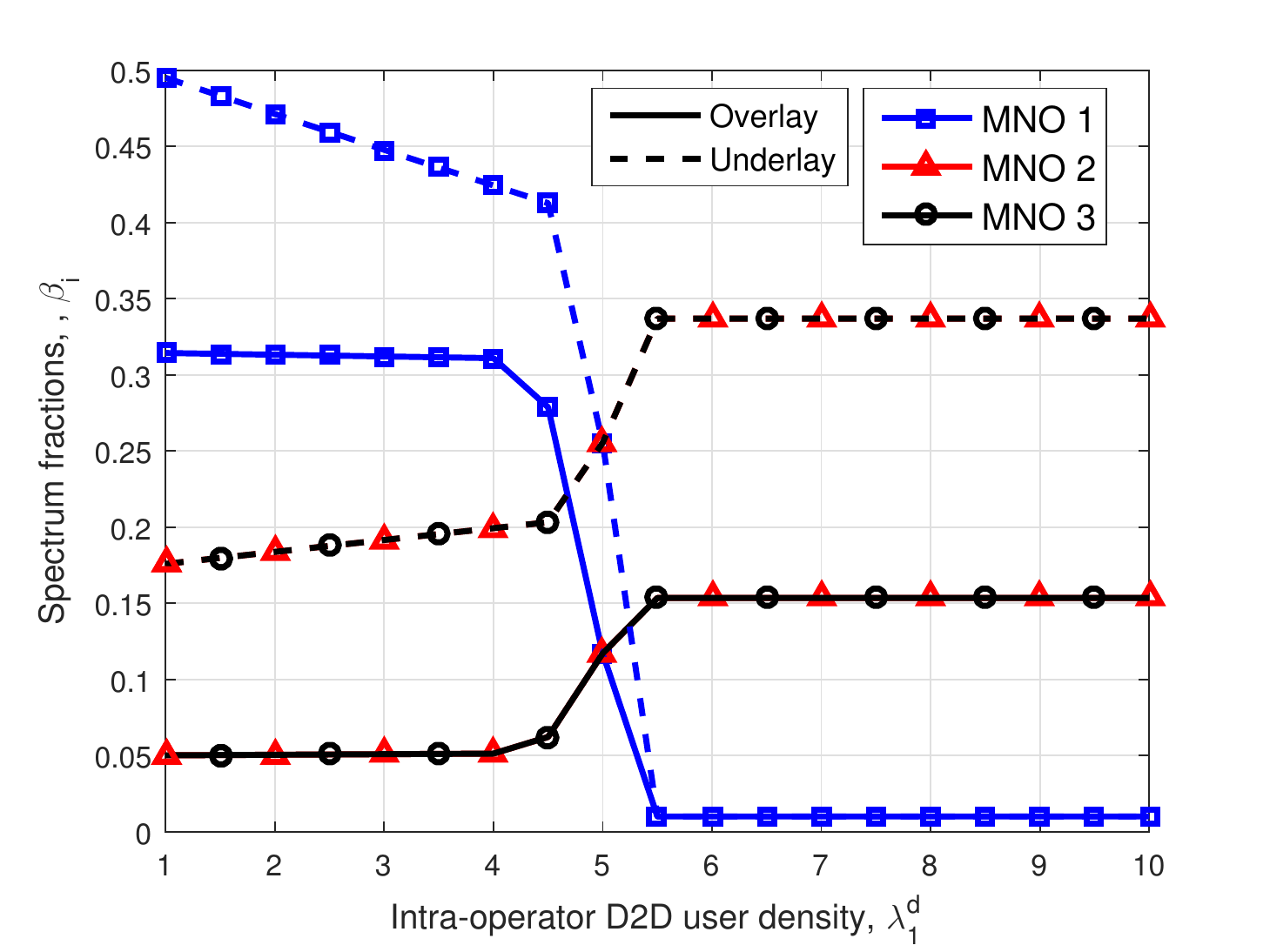}} & 
\hspace{-18pt}
\subfigure[\label{Numerical:fig:asym:b}]
{\includegraphics[width=0.335\textwidth]{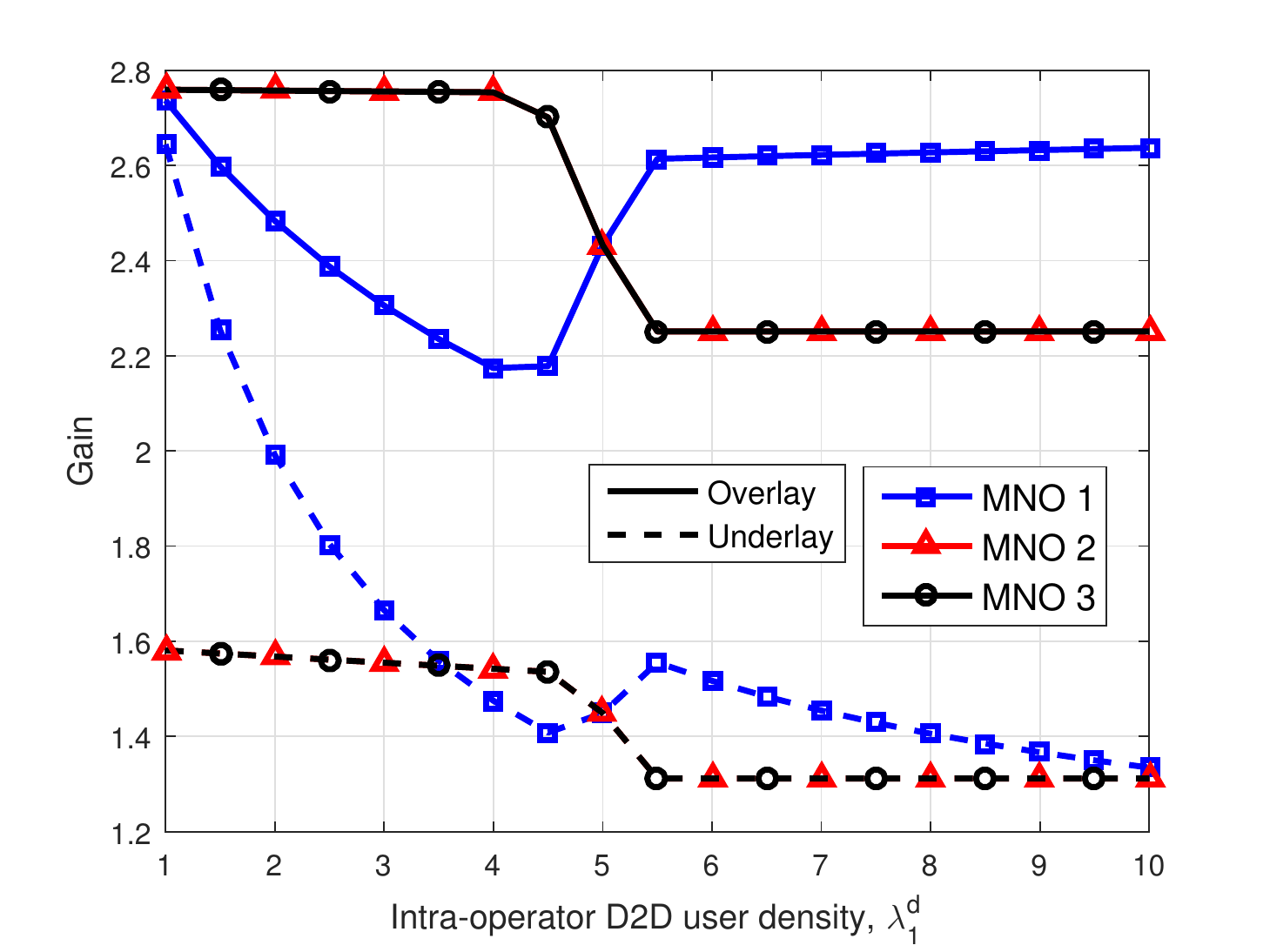}} & 
\hspace{-18pt}
\subfigure[\label{Numerical:fig:asym:c}]
{\includegraphics[width=0.335\textwidth]{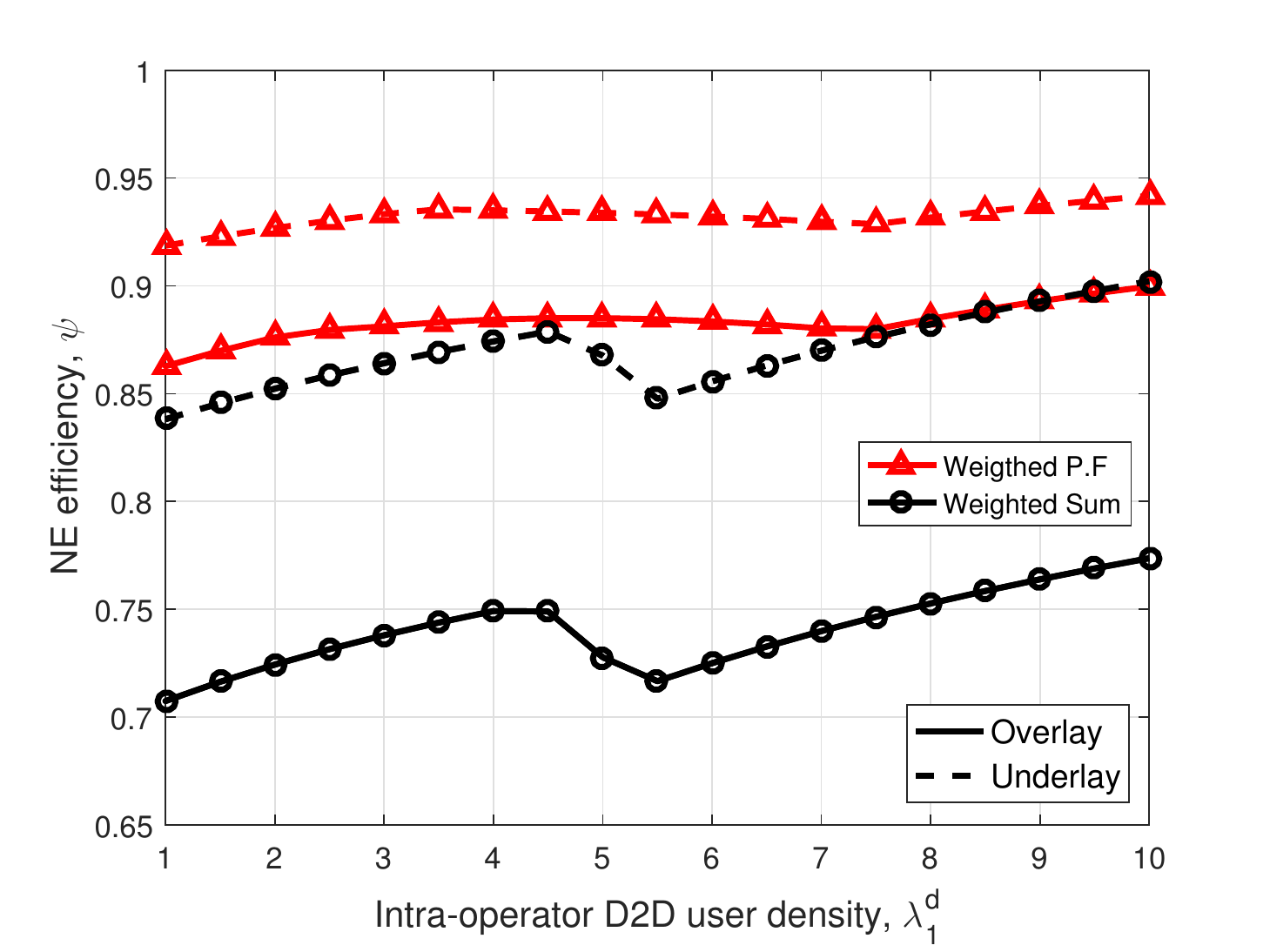}}\\ 
\end{tabular}
\caption{(a) Spectrum fraction, $\beta_i$, (b) Performance gain, and (c) NE efficiency, w.r.t $\lambda_1^d$, when $N =3$ and $\lambda_{i\geq2}^d = 5$.}
\end{figure*}

\subsection{Performance}
We evaluate the performance of the \acp{OP} for different intra-D2D densities for \ac{OP} 1, $\lambda_1^d$, while we fix $\lambda_2^d$ and $\lambda_3^d$ to be equal to the cellular user density. We depict the spectrum fractions contributed by the \acp{OP} in Fig.~\ref{Numerical:fig:asym:a}, and assess the sum rate gain for the \acp{OP} as compared to the case without inter-D2D support in Fig.~\ref{Numerical:fig:asym:b}. The gain is computed as follows: $\frac{w_i^c\cdot Q_i^c + w_i^d\cdot Q_i^d  + w_i^s\cdot Q_i^s}{(w_i^c+w_i^s)\cdot Q_{i,o}^{c} + w_i^d\cdot Q_{i,o}^{d}}$ where $Q_{i,o}^c$ and $Q_{i,o}^d$ are the average rates of cellular users and intra-D2D users for the case without inter-D2D support, evaluated after setting the user fraction in inter-D2D mode $q = 0$  and the D2D spectrum allocation factor $\beta = 0$, with the D2D spectrum allocation factor $\beta_i^d$ given the constraints $\tau_i^c$ and $\tau_i^d$. In general, asymmetric \acp{OP} would contribute unequal amounts of spectrum. All \acp{OP} experience performance gain. One can see that symmetric \acp{OP} achieve around $145\%$ gain.

 For densities $\lambda_1^d<5$, \ac{OP} 1 who has less network load contributes the higher fraction of spectrum to the spectrum pool, see Fig.~\ref{Numerical:fig:asym:a}, since it has enough capacity to satisfy its own constraints. Hence, the opponents enjoy more benefit from \ac{CoPSS} than \ac{OP} 1 does, see Fig.~\ref{Numerical:fig:asym:b}. The inter-D2D user rate is larger than intra-D2D user rate, $Q_i^s > Q_i^d$, since the total amount of spectrum committed to the shared band is larger than the one allocated to intra-D2D users in D2D mode, $\beta > \beta_i^d$. However, due to the decreasing fraction of inter-D2D users, $w_i^s$, the gain obtained by \ac{OP} 1 decreases along with the densities $\lambda_1^d<5$. Meanwhile, for density around $\lambda_1^d \!= \!5$, all \acp{OP} come to have same network load and thus \acp{OP} 2 and 3 who benefited by contributing less spectrum fractions contribute more, and \ac{OP} 1 contributes less spectrum. All \acp{OP} achieve equal gains at $\lambda_1^d \!=\! 5$. On the other hand, for densities $\lambda_1^d\!>\!5$, \ac{OP} 1 contributes only a small fraction for the signaling channel. Other \acp{OP} still benefit by contributing some fraction, i.e., $\beta_{i>1} \!=\! 0.15~\mbox{and}~0.34$ in overlay and underlay, respectively.

One can see that \acp{OP} using the underlay principle would contribute more spectrum for inter-D2D communications, and experience less performance gain, under the same constraint target values with the intra-D2D overlay approach. While cellular and intra-D2D users suffer from the mutual interference and less average spectral efficiency, the underlay scheme appears to achieve higher rates for cellular and intra-D2D users due to wider spectrum allocation. This enables an \ac{OP} to have enough capacity satisfying its own constraints. Thus, an \ac{OP} contributes more spectrum and achieves less gain compared to the no spectrum sharing case where a higher rate is already obtained in underlay compared to the overlay scenario.

Fig.~\ref{Numerical:fig:asym:c} shows the efficiency of the non-cooperative solution, the ratio of the utility sum at the NE point obtained by Algorithm 1 compared to the utility sum at the pareto-optimal point, $\psi$. A socially optimal solution can be viewed as one type of fair allocation. The weighted \ac{P.F} utility function  allocates resource more fairly between two groups, e.g., intra-operator users and inter-operator users. An \ac{OP} with this utility, who might be able to increase its own utility by allocating spectrum to the intra-D2D users or to the inter-D2D users, tends to avoid choosing an extreme value in his own strategy space for preference maximization. Such a rational manner in P.F utility function brings more fairness among \acp{OP}, and thus yields a result closer to the optimal one. The obtained NE efficiency can be used as a lower bound for a two-operator non-cooperative spectrum sharing game, since increasing the number of \acp{OP} leads to the NE efficiency compromise.

\section{Conclusions}
We studied spectrum allocation for D2D communication considering different mobile network \acp{OP}. We modeled the interactions between \acp{OP} as a non-cooperative game. We showed that the formulated game has a unique NE if every \ac{OP} has a concave utility on the box-constrained region and all eigenvalues of derivatives of iterative response process are less than unity. Uniqueness can be identified in a distributed manner. The non-cooperative algorithm based on the \ac{OP}'s best response might not converge to the NE due to myopically overreacting to the responses of the other \acp{OP}. To resolve this instability, we proposed a \ac{JP} strategy update algorithm with a proper smoothing parameter. Using the \ac{JP} update we were able to study the system and draw useful remarks. Asymmetric \acp{OP} contribute an unequal amount of spectrum for D2D support. An \ac{OP} may contribute a small amount of spectrum, but still the opponents may have the incentive to contribute more due to the D2D proximity gain. We illustrated that participating \acp{OP} may experience significant performance gains depending on the operator-specific network load, utility and design constraints.

\appendix
\subsection{Proof of Proposition \ref{numerical:concave}\label{ref:numerical:concave}}
For intra-D2D overlay in the weighted sum rate utility, we show 
$\frac{\partial^2 U_i}{{\partial\beta_i}^2} = w_i^d {Q_i^d}{''} \!+\! w_i^s {Q_i^s}{''}  = (W^c \beta_i^c R_i^c){''} \!+\! (W^d \beta_i^d R_i^d){''} \!+\! (W^s \beta R^s){''} \!<\!0$ where $W_i^c \!=\! w_i^d (1-\delta_i) \!+\! w_i^s(1-q)$, $W_i^d \!=\! w_i^d \delta_i$, and $W_i^s \!=\! w_i^s q$, which narrows down to $\frac{\partial^2 U_i}{{\partial\beta_i}^2} \!= \!(W^d \beta_i^d R_i^d){''} \!+ \!(W^s \beta R^s){''} \!<\!0$, by using the coverage probability for the cellular uplink, $R_i^c$, derived and verified in~\cite{Cho2014} where the cellular system is interference-limited. The constraint in (\ref{eq:P1:b}) affects the upper limit of $\beta_i$ for a fixed $\delta_i$, proven in Prop. \ref{numerical:constraint}. The maximums of the D2D user rates, $Q_i^d$ and $Q_i^s$ are along the border of the feasibility region, i.e., $\beta_i^d \!=\! 1\!-\! \beta_i\!-\!\frac{\tau_i^c}{R_i^c}\! >\! 1\!-\! \beta_i \!-\!\beta_i^c$. 
By using the Leibniz rule~\cite{Abramowitz} and $\nu_i^d = \nu^s = 1$, both terms are negative, $(\beta_i^d R_i^d){''} <0$ and $(\beta R^s){''} <0$, thus ${Q_i^d}{''}\!<\!0$ and ${Q_i^s}{''}\!<\!0$, proven in \cite{Cho2015}. For intra-D2D underlay in the weighted sum rate utility, $\beta_i^c$ and $\beta_i^d$ are replaced by $\beta_i^{cd}$, yielding $\frac{\partial^2 U_i}{{\partial\beta_i}^2} \!=\! (W^d \beta_i^{cd} R_i^d){''} + (W^s \beta R^s){''} \!<\!0$. In a similar manner, we have $(\beta_i^{cd} R_i^d){''} <0$. In the P.F rate utility, $\frac{\partial^2 U_i}{{\partial\beta_i}^2} \!=\! \frac{w_i^d\{{Q_i^d}\cdot{Q_i^d}{''} - ({Q_i^d}{'})^2 \}}{{Q_i^d}^2}+  \frac{w_i^s\{{Q_i^s}\cdot{Q_i^s}{''} - ({Q_i^s}{'})^2 \}}{{Q_i^s}^2} \!<\!0$, holds true, due to ${Q_i^d}{''}<0$ and ${Q_i^s}{''}<0$.

\subsection{Proof of Proposition \ref{numerical:constraint}\label{ref:numerical:constraint}}
We show that $h_i^d$ and $h_i^c$ are increasing i) in $\beta_i^{cd}$, $\beta_i^{d}$ and $\beta_i^c$, and ii) in $\delta_i$. i) For a fixed $\delta_i$, $h_i^d$ is increasing in $\beta_i^d$ and in $\beta_i^{cd}$, respectively. We show $\frac{\partial h_i^d}{{\partial\beta_i^d}} >0$. By using the Leibniz rule~\cite{Abramowitz}, and $\nu_i^d = 1$, it is sufficient to show ${\partial h_i^d}/{{\partial\beta_i^d}}\overset{(p1)}{\geq} \!\! \int_0^{\infty} \frac{\delta_i e^{-{\gamma {\beta_i^d}}/{\eta} - \gamma \delta_i\overline{c}}(1-\frac{{\beta_i^d}\gamma}{\eta})}{1+\gamma}   \mbox{d}\gamma  \nonumber \!\overset{(p2)}{=} \frac{- \delta_i\rho}{\rho+ \delta_i\overline{c}} + \frac{\delta_i(\rho+1){E}_1(\rho+\delta_i\overline{c})}{e^{-\rho - \delta_i\overline{c}}}\nonumber 
\overset{(p3)}{>} \delta_i^2\overline{c} >0\nonumber
$ where $\eta = \frac{P_d l(d)}{\sigma^2}$, inequality ($p1$) holds true due to $\int_{0}^{\infty} \frac{2\pi \delta_i \lambda_i^d \gamma l(r) r}{l(d) \!+\! \gamma\cdot l(r) } {d}r \!<\!\gamma \delta_i \overline{c}$ and $\overline{c} = 2\pi \lambda_i^d \int\nolimits_{0}^{\infty} \frac{l(r) r }{l(d)} {d}r$. Equality ($p2$)  holds true due to $\int_0^{\infty} \frac{e^{- (\rho+c) x} (1-\rho x)}{(1+x)}  ~dx \!\!=\!\!  \frac{-\rho}{\rho+c} \! +\! \frac{(\rho+1) {E}_1(\rho+c)}{e^{-\rho-c}}$ where $\rho = \frac{\beta_i^d}{\eta}$, $c = \delta_i \overline{c}$, and $E_1(x)$ is the exponential integral. For $\rho + \delta_i c > 0$, inequality ($p3$)  holds true due to a continued fraction form of $E_1(x)$ larger than $\frac{e^{-x}}{1+x}$ from \cite[5.1.22]{Abramowitz}. Thus, $h_i^d$ is increasing in $\beta_i^d$ for the intra-D2D overlay, and also in $\beta_i^{cd}$ for the intra-D2D underlay after replacing $\beta_i^d$ by $\beta_i^{cd}$. Since $h_i^c$ is linearly increasing in $\beta_i^c$, $\beta_i^{c,min}$,$\beta_i^{d,min}$, and $\beta_i^{cd,min}$ determine $\beta_i^{max}$.

ii) For a fixed $\beta_i$, $h_i^d$ is increasing in $\delta_i$ for both intra-D2D underlay and overlay. We show $\frac{\partial h_i^d}{{\partial\delta_i}}>0$. In a similar manner to i) above, it is sufficient to show $\frac{\partial h_i^d}{{\partial\delta_i}} \!=\! \beta_i^d \int_0^{\infty} \frac{\mathcal{P}_i^d \left(1\!-\!\delta_i (C_1'\!+\!C_2') \right)}{1+\gamma}   \mbox{d}\gamma\!\! \overset{(p4)}{\geq}\!\! \beta_i^d \int_0^{\infty} \!\frac{\mathcal{P}_i^d \left(1\!-\! \delta_i C_1'\!\right)}{1+\gamma}  \mbox{d}\gamma 
\!\!\!\overset{(p5)}{>} \beta_i^d(\beta_i^d/\eta) >0\nonumber$
where $C_1' \!=\! \!\int_{0}^{\infty} \frac{ 2\pi \lambda_i^d \gamma l(r) r}{l(d) + \gamma\cdot l(r) } {d}r$ and $C_2' =  \int_{0}^{\infty} \frac{\alpha_i' 2\pi\lambda_i^b \gamma l(r) r}{l(d)P_d/P_c + \gamma l_d(r)} {d}r$ for the intra-D2D underlay, and $C_2' = 0$ for the intra-D2D overlay. Note that the probability a BS is active, $\alpha_i$\cite{Cho2014}, decreases in $\delta_i$, since less D2D users select in cellular mode. Thus $\alpha_i'$ is negative and $C_2'$ is non-positive. Inequality ($p4$) holds true due to $1-\delta C_1' - \delta C_2' \geq 1-\delta C_1'$ for both intra-D2D underlay and overlay. Inequality ($p5$) holds true due to $C_1' <  \gamma  \overline{c}_1 $ in $\mathcal{P}_i^d$ where $\overline{c}_1 = \!\int_{0}^{\infty} { 2\pi  \lambda_i^d l(r) r}/{l(d)}{d}r$, and due to the relation in ($p2$) where $\rho = \delta_i \overline{c}_1$ and $c = \beta_i^d/\eta$. Thus, $h_i^d$ is increasing in $\delta_i$ for both intra-D2D underlay and overlay. For a fixed $\beta_i$, $h_i^c$ is increasing in $\delta_i$ for the intra-D2D overlay. We show ${\partial h_i^c}/{{\partial\delta_i}} \!=\! \int_0^{\infty} {\beta_i^c \mathcal{P}_i^c \left({\nu_i^{c}}' - {\nu_i^{c}}(C_3'\!+\!C_4') \right)}/({1+\gamma})   \mbox{d}\gamma\!\! >0\nonumber$. Note that the portion of time a user in cellular mode is active in the uplink, $\nu_i^c$ \cite{Cho2014}, increases in $\delta_i$, and ${\nu_i^{c}}'$ is positive, since less D2D users select in cellular mode. Thus, we show $\int_0^{\infty} {\beta_i^c \mathcal{P}_i^c \left(-C_3'\!-\!C_4' \right)}/({1+\gamma})\mbox{d}\gamma\!\! >0\nonumber$. We have $C_3'= \int_{d}^{\infty} {\alpha_i' 2\pi\lambda_i^b \gamma l(r) r}/({l(d) + \gamma l(r)}) {d}r<0$ due to $\alpha_i'<0$. Thus, the inequality above holds with $C_4' = 0$ for the intra-D2D overlay. However, for the intra-D2D underlay, the inequality above does not hold yet due to $C_4' \!=\! \!\int_{0}^{\infty} {2\pi \lambda_i^d \gamma l(r) r}/({l(d)P_c/P_d  + \gamma\cdot l(r)}) {d}r \geq 0$. Instead, as discussed in Section~\ref{sec:System_model}, we identify $\tau_i^d$ yielding $\beta_{i,d}^{cd} > \beta_{i,c}^{cd}$. To this, the constraint in~ \eqref{eq:P1:b} is strictly satisfied with $\beta_{i,d}^{cd}$ or  the constraint in~\eqref{eq:P1:c} is violated with $\beta_{i,c}^{cd}$, i.e., $\tau_i^d > \delta_i \beta_{i}^{d} R_i^d|_{\beta_i^d  = \tau_i^c/R_i^c}$. 
To sum up, $h_i^d$ is, in the intra-D2D overlay, increasing in $\beta_i^{d}$ and in $\delta_i$, and also, in the intra-D2D underlay, increasing in $\beta_i^{cd}$ and in $\delta_i$. $h_i^c$ is, in the intra-D2D overlay, is increasing in $\beta_i^c$ and in $\delta$, and also, in the intra-D2D underlay, increasing in $\beta_i^{cd}$. And with ${\tau_i^d} > {\tau_i^c}{R_i^d }/{R_i^c}{|_{\beta_i^d = 0}}$, the constraint set has ascending property.

\subsection{Proof of Lemma \ref{lemma_rowdiag}\label{ref:lemma_rowdiag}}
We consider the row diagonally dominant matrix where the diagonal element in each row of $-H(\boldsymbol{\beta}')$ exceeds the sum of the moduli of the off-diagonal element. Then, in each row of $\!-H(\boldsymbol{\beta}')$,  we have $|\!-\!H_{ii}(\boldsymbol{\beta}')|\!>\! \sum_{j\neq i, j\in \mathcal{N}}|\!-\!H_{ij}(\boldsymbol{\beta}')| \!\geq\! \sum_{j\neq i, j\in {\mathcal{\acute{N}}}}|-\!H_{ij}(\boldsymbol{\beta}')| \geq \sum_{j\neq i, j\in {\mathcal{\bar{N}}}}|-\!H_{ij}(\boldsymbol{\beta}')|, \forall i$. Thus, if $\!-\!H(\boldsymbol{\beta}')$ is row diagonally dominant with positive diagonal elements, arbitrary principal sub-matrices, $-H(\boldsymbol{\beta}')|_{\mathcal{\bar{N}}}$ for any ${\mathcal{\bar{N}}}\subseteq {\mathcal{\acute{N}}}$ are positive row diagonally dominant, and have positive determinants \cite{Ortega}. That is, $-H(\boldsymbol{\beta}')$ is $P$-matrix, and $\boldsymbol{\beta}'$ always has a positive one index, $Ind(\boldsymbol{\beta}')= sign(det(-H(\boldsymbol{\beta}'))) = 1$, no matter whether $\boldsymbol{\beta}'$ is interior or boundary NE.

\subsection{Proof of Proposition \ref{prop_Hfun}\label{ref:prop_Hfun}}
The mapping function $M$ is typically defined implicitly through the first-order conditions\footnote{It only tells us about the behavior of $M$ near the point $\beta$}. The first order conditions $\forall i, {\partial U_i(M_i,\boldsymbol{\beta_{-i}})}/{\partial \beta_i}=0$ hold for all $\boldsymbol{\beta_{-i}}$ \cite[p25]{Cachon}. For $j\neq i$ we differentiate the first order condition for $i$ with respect to $\beta_j$, then $\frac{\partial^2 U_i}{\partial \beta_i^2}\frac{\partial M_i}{\partial \beta_j} \!+\! \frac{\partial^2 U_i}{\partial \beta_i\partial \beta_j}\!=\!0$,  yielding
\begin{equation} 
\begin{array}{llll} \label{eq:implicit_func}
\frac{\partial M_i(\beta_j)}{\partial \beta_j} \!\!=\!\! -\frac{\partial^2 U_i(M_i(\boldsymbol{\beta_{-i}}),\boldsymbol{\beta_{-i}})}{\partial \beta_i\partial \beta_j} \!\!\left(\frac{\partial^2 U_i(M_i(\boldsymbol{\beta_{-i}}),\boldsymbol{\beta_{-i}})}{\partial \beta_i^2}\right)^{\!\!-1}\!.
\end{array}
\end{equation}
The Hessian matrix is defined as $H(\boldsymbol{\beta}')\!\! = \!\![\nabla u_1\nabla u_2 \cdots\nabla u_N]^T$ where $\nabla u_i = [\frac{\partial^2 U_i}{\partial \beta_1 \partial \beta_i}\frac{\partial^2 U_i}{\partial \beta_2 \partial \beta_i}\cdots\frac{\partial^2 U_i}{\partial \beta_N \partial \beta_i}]$ is $i$-th row of $H(\boldsymbol{\beta}')$ and can also be expressed as a multiplication of two $N \times N$ matrics, $H \!=\! H^U  H^T$. The matrix, $H^U \!\!=\! [H_{ij}^U]_{\forall i,j}$ has elements with $H_{ij}^U \!=\!\frac{\partial^2 U_i}{\partial\beta_j^2}$ for $i\!=\!j$ and $H_{ij}^U \!=\!0$ for $i\neq j$. The matrix, $H^T \!\!=\! [H_{ij}^T]_{\forall i,j}$ has elements with $H_{ij}^T \!=\!1$ for $i\!=\!j$ and $H_{ij}^T \!=\! -\frac{\partial M_i}{\partial\beta_j}$ for $i \!\neq\! j$.

\subsection{Proof of Lemma~\ref{lemma_index5}\label{ref:lemma_index5}}
The condition for the contraction mapping, $\!\!\|T\|_{\infty}\!\!=\!\!\|M'(\boldsymbol{\beta}')\|_{\infty}\! \!<\!1\!$ ({\color{black}Lemma \ref{lemma:contract2}}), is equal to $\sum_{j}|{\frac{\partial M_i(\boldsymbol{\beta}')}{\partial\beta_j}}| \!\!<\!\!1, \!\forall i$ which can be expressed as, according to Eq. (\ref{eq:implicit_func}), $\sum_{j}|{\frac{\partial M_i(\boldsymbol{\beta}')}{\partial\beta_j}}|  = \sum_{j\neq i}|-{\frac{\partial^2 U_i(\boldsymbol{\beta}')}{\partial\beta_i \beta_j}}/{\frac{\partial^2 U_i(\boldsymbol{\beta}')}{\partial\beta_i^2}}| = \sum_{j\neq i}\frac{|-H_{ij}(\boldsymbol{\beta}')|}{|-H_{ii}(\boldsymbol{\beta}')|}\!<\!1, \!\forall i$ and thus satisfies a row diagonally dominant matrix, $|-H_{ii}(\boldsymbol{\beta}')|> \sum_{j\neq i}|-H_{ij}(\boldsymbol{\beta}')|, \forall i$. If $-H(\boldsymbol{\beta}')$ is row diagonally dominant with positive diagonal elements ensured by the concavity of the utility, $\boldsymbol{\beta}'$ has a positive one index (Lemma~\ref{lemma_rowdiag}).

\subsection{Proof of Proposition \ref{prop:JC}\label{ref:prop:JC}}
According to {\color{black}Lemma \ref{lemma:contract2}}, the condition based on the infinity norm is sufficiently satisfied if each absolute row sum of the jacobian matrix based on Jacobi update is less than one, $\sum_{j}|J_{ij}^{(t)}| <1, \,\,\,\,\forall i$ where $\sum_{j=1}^{n}|J_{ij}^{(t)}| = |1-\kappa_i^{(t)}| + \kappa_i^{(t)} \cdot \sum_{j\neq i} |J_{ij}^{BR(t)}|$.  The condition $\sum_{j}|J_{ij}^{(t)}| <1$ implies that $\kappa_i^{(t)}( \sum_{j\neq i} |J_{ij}^{BR(t)}|-1)\!<\!0$ for $\kappa_i^{(t)} \leq 1$ and  $\kappa_i^{(t)} ( \sum_{j\neq i} |J_{ij}^{BR(t)}|+1)\!<\!2$ for $\kappa_i^{(t)} > 1$. Therefore, if $\sum_{j\neq i} |J_{ij}^{BR(t)}|\!<\!1$, $\sum_{j}|J_{ij}^{(t)}| <1$ is satisfied for $0 < \kappa_i^{(t)}\!<\! {2}/({\sum_{j\neq i} |J_{ij}^{BR(t)}|+1})$ \cite[p.537]{Basar}, due to $0 < \kappa_i^{(t)}\leq 1$ for $\kappa_i^{(t)} \leq 1$ and $1 < \kappa_i^{(t)}\!<\! {2}/({\sum_{j\neq i} |J_{ij}^{BR(t)}|+1})$ for $\kappa_i^{(t)} > 1$.

\subsection{Proof of Lemma~\ref{lemma:jc:dominant}\label{ref:lemma:jc:dominant}}
The conditions for the lower and upper limits of the Gerschgorin's circle region to be in absolute value less than one can be expressed in terms of $\kappa_i^{(t)}$ as $\kappa_i^{(t)}({\sum_{j\neq i} |J_{ij}^{BR(t)}|}+1) < {2}$ and $\kappa_i^{(t)}(\sum_{j\neq i} |J_{ij}^{BR(t)}|-1)<0$, respectively. We observe that the condition for the upper limit is $\kappa_i^{(t)}\!>\!0$  if $\sum_{j\neq i} |J_{ij}^{BR(t)}|\!<\!1$, while the condition for the lower limit is $\kappa_i^{(t)} \!<\! {2}/({\sum_{j\neq i} |J_{ij}^{BR(t)}|}+1)$. Since the maximum eigenvalue is less than one, instability of ${J}^{(t)}$ can only be caused by the minimum eigenvalue. If the minimum eigenvalue is less than $-1$, the \ac{BR} is not stable according to {\color{black}Remark \ref{remark:3}}. Satisfying the condition for the lower limit of the circle region, $\kappa_i^{(t)}\!\!<\!\!{2}/({\sum_{j\neq i}\! |J_{ij}^{BR(t)}|}+1)$ sufficiently ensures the convergence of the \ac{JP} scheme.

\subsection{Proof of Lemma~\ref{prop:pooling}\label{ref:prop:pooling}}
The characteristic polynomial of ${J}$ can be written as $G(\xi)= det(\xi I-{J}) = det(\xi I - {\Delta} - \Omega
\cdot \boldsymbol{1}^T)$ where ${\Delta}$ is $N\times N$ diagonal matrix, $diag[J_{11}\!-\!J_1,\!~\cdots, \!\!~ J_{NN}\!-\!J_N]$, $\Omega$ is $N\times 1$ column vector, $[J_{1},\cdots,\!J_N]^T$, and $\boldsymbol{1}^T$ is $1\!\times\! N$ row vector. Using simple algebra, $G(\xi)= det(\xi I-{J}) = det(\xi I - {\Delta} - \Omega
\cdot \boldsymbol{1}^T) = det(\xi I - {\Delta})\cdot det(I - (\xi I - {\Delta})^{-1} \Omega \cdot\boldsymbol{1}^T)\overset{(p1)}{=} det(\xi I - {\Delta})\cdot (1 -  \boldsymbol{1}^T(\xi I - {\Delta})^{-1} {\Omega})$ where $(\xi I - {\Delta})^{-1}$ is a diagonal matrix due to the fact that inverse of a diagonal matrix is diagonal. The equality (p1) holds because for all $N$-element real column vectors, $\boldsymbol{x}$ and $\boldsymbol{y}$, we have $det(I + \boldsymbol{x}\cdot \boldsymbol{y}^T) = 1 +  \boldsymbol{y}^T\cdot\boldsymbol{x}$. Therefore, we have $det(I- (\xi I -{\Delta})^{-1} {\Omega}\cdot \boldsymbol{1}^T) = 1 + g(\xi)$ where $g(\xi) = - \boldsymbol{1}^T\cdot (\xi I - {\Delta})^{-1} {\Omega}$, and thus $G(\xi) = (1+g(\xi))  \prod_{i} (\xi- d_i)  = 0$ where  $d_i = J_{ii}-J_i^{}$ and $g(\xi) = \sum_{i}\frac{-J_i^{}}{\xi-d_i}$ noting that $(\xi I - {\Delta})^{-1} \Omega$ is a column vector. Without loss of generality, we could let $\xi_1 \leq \!\!\cdots\!\! \leq \xi_N$ and $d_1 \leq \!\!\cdots \!\!\leq d_N$. If $G(\xi)$ is a continuous polynomial, then the eigenvalues of $J$, $\xi_1,\cdots,\xi_N$, are the roots of $G(\xi) = 0$. There exists a root $\xi_i$, if there exists two numbers $\xi_i^{min}$ and $\xi_i^{max}$, $\xi_i^{min} < \xi_i < \xi_i^{max}$ such that $G(\xi)$ evaluated at $\xi_i^{min}$ and $\xi_i^{max}$ shows opposite signs due to the continuous property of $G(\xi)$. Note that $g(\xi)= -1$ can be evaluated at $\xi \neq d_i, \forall i$, and the roots of $g(\xi)= -1$ are identical to the roots of $G(\xi) = 0$. $g(\xi)$ is continuous in the following intervals $(-\infty, d_1),(d_1, d_2), \cdots, (d_N, +\infty)$, and it has $\lim_{\xi\rightarrow -\infty} g(\xi) = \lim_{\xi\rightarrow +\infty} g(\xi) = 0$. When $J_i < 0$, $g(\xi)$ decreases due to ${\partial g(\xi)}/{\partial \xi} = \sum_i {J_i}/{(\xi-d_i)^2}<0$. And it has $\lim_{\xi\rightarrow d_i^{-}} g(\xi) = -\infty$, $\lim_{\xi\rightarrow d_i^{+}} g(\xi) = +\infty, \forall i$. $g(\xi_N)$ is positive for $d_N < \xi_N < \infty$ due to the continuity property and the $\lim_{\xi\rightarrow +\infty} g(\xi) = 0$. Hence, the roots of $g(\xi)= 0$ are in $(-\infty, d_1), \cdots, (d_{N-1}, d_{N})$, thus $-\infty< \xi_1< d_1$ and $d_{i-1}< \xi_i< d_i$ where $2\leq i \leq N$. For $i\neq 1$, $|\xi_i| < 1$ is guaranteed if $-1 < d_1$ and $d_N < 1$, because $\xi_i$ where $2\leq i\leq N$ is between $d_1$ and $d_N$, which can be ensured by $|d_i|< 1$. For $i=1$, $|\xi_1|<1$ is guaranteed if $-1 < \xi_1 < d_1$, because $\xi_1$ is between $-\infty$ and $d_1$. The condition can be ensured by $g(-1)> -1$, because $g(\xi)$ is decreasing in $(-\infty, d_1)$ and $g(\xi_1) = -1$ since $\xi_1$ is a root of $1+g(\xi)=0$. Hence, the condition for $-1< \xi_1 < d_1$ is $g(-1) = \sum_i {J_i}/(1+d_i) \!>\! -1$.

\subsection{Proof of Proposition~\ref{lemma:pooling2}\label{ref:lemma:pooling2}}
Following the proof of {\color{black}Lemma \ref{prop:pooling}}, all of the eigenvalues in ${J}^{BR}$ are less than one, if $J_{i}^{} = J_{ij}^{BR}< 0$ and $d_N = J_{ii}-J_i^{} = -J_{ij}^{BR} < 1, \forall i$, implying that there is a unique NE since the maximum eigenvalue is less than a positive one ({\color{black}Remark \ref{remark:3}}).

\subsection{Proof of Proposition \ref{lemma:pooling3} \label{ref:lemma:pooling3}}
According to {\color{black}Lemma \ref{prop:pooling}}, all of the eigenvalues in ${J}^{BR}$ are inside the unit circle of the complex plain, if (i) $J_{i}^{}=J_{ij}^{BR}< 0$, (ii) $|J_{ii}-J_i^{}|= |-J_{ij}^{BR}| < 1, \forall i$, and (iii) $\sum_i (-J_i^{})/{(1+J_{ii}-J_i^{})} = \sum_i ({-J_{ij}^{BR}})/{(1-J_{ij}^{BR})} <1$ satisfied if each \ac{OP} satisfies $({-J_{ij}^{BR}})/{(1-J_{ij}^{BR})}<{1}/{N}$.

 \subsection{Proof of Proposition~\ref{lemma:pooling4}\label{ref:lemma:pooling4}}
According to {\color{black}Lemma \ref{prop:pooling}}, all of the eigenvalues in ${J}^{(t)}$ are inside the unit disk, if (i) $J_{i}^{}=  \kappa_i^{(t)} J_{ij}^{BR}< 0$, (ii) $|J_{ii}-J_i^{}|= |1-\kappa_i^{(t)}- \kappa_i^{(t)} J_{ij}^{BR}| < 1, \forall i$, and (iii) $\sum_i (-J_i^{})/{(1+J_{ii}-J_i^{})} = \sum_i ({-\kappa_i^{(t)}  J_{ij}^{BR}})/{(2-\kappa_i^{(t)}- \kappa_i^{(t)} J_{ij}^{BR})} <1$. For (ii) above, we have $-\kappa_i^{(t)}(J_{ij}^{BR}+1)<0$ and $-\kappa_i^{(t)}(J_{ij}^{BR}+1)>-2$. If $J_{ij}^{BR}<-1$, then $\kappa_i^{(t)} <0$. Thus, it should be $J_{ij}^{BR}>-1$. This results in the following range, $0<\kappa_i^{(t)}<{2}/(1+{J_{ij}^{BR}})$. The condition (iii) above is satisfied if each \ac{OP} satisfies the following condition ${-\kappa_i^{(t)}  J_{ij}^{BR}}/{(2-\kappa_i^{(t)}- \kappa_i^{(t)} J_{ij}^{BR})}< {1}/{N}$, equivalently $\kappa_{i}^{(t)} <{2}/({1- (N-1) J_{ij}^{BR}})$. Due to $J_{ij}^{BR}<0$, $\kappa_i^{(t)}< {2}/({1+ (N-1) |J_{ij}^{BR}|})$. To sum up, we have $-1\!<\!J_{ij}^{BR}\!<\!0$, and $0\!<\!\kappa_i^{(t)}\!<\!\kappa_{i,max}^{(t)}$ where $\kappa_{i,max}^{(t)} \!=\!{2}/({1+(N-1)|J_{ij}^{BR}|})$. When $-1<J_{ij}^{BR}\leq -1/(N-1)$, the \ac{BR} will not converge and $\kappa_{i,max}^{(t)}$ becomes less than one.

\subsection{Proof of Proposition \ref{numerical:uniqueness2} \label{ref:numerical:uniqueness2}}
According to equation (\ref{eq:implicit_func}), the sufficient condition in Prop. \ref{lemma:pooling2} can be expressed as $-1 \!<\! J_{ij}^{BR}\!=\!  -{\partial_{ij}^2 U_i}/{\partial_{ii}^2 U_i}\!<\!0$ where $\partial_{xy}^2 U_i \!=\! {\partial^2 U_i}/({\partial \beta_x \partial \beta_y})$. Due to the utility structure subject to the system framework in \ac{CoPSS} scenario: in-band overlay spectrum allocation, i.e., $\beta_i^{cd} + \beta_i = 1$, and a shared spectrum pool usage, i.e., $\beta = \sum_i \beta_i$, we have $\partial_{ij}^2 U_i = w_i^s  \partial_{ij}^2 U_i^{s} = w_i^s  \partial_{ii}^2 U_i^{s}$ and $\partial_{ii}^2 U_i = w_i^s  \partial_{ii}^2 U_i^{s}+ (1-w_i^s) \partial_{ii}^2 U_i^{d}$, where $U_{i}^{s}$ and $U_{i}^{d}$ are the performances for inter-D2D users and intra-D2D users, i.e., $U_{i}^{k} = Q_i^k$ or $U_{i}^{k} = \log(Q_i^k)$ for $k$-type users, yielding $J_{ij}^{BR} = - \{w_i^s  \partial_{ii}^2 U_i^{s} \} /\{w_i^s  \partial_{ii}^2 U_i^{s}+ (1-w_i^s) \partial_{ii}^2 U_i^{d}\}$. Thus, the sufficient condition is satisfied if i) $|w_i^s  {\partial_{ii}^2 U_i^{s}} + (1-w_i^s)  \partial_{ii}^2 U_i^{d}| > |w_i^s {\partial_{ii}^2 U_i^{s}}|$ for $0 \leq w_i^s < 1$, and ii) $\mbox{sign}\{w_i^s  {\partial_{ii}^2 U_i^{s}} + (1-w_i^s){\partial_{ii}^2 U_i^{d}}\} = \mbox{sign}\{w_i^s  {\partial_{ii}^2 U_i^{s}}\}$ for $0 <w_i^s \leq 1$, which are met by the concavity of $U_i$ for $0 \leq w_i^s \leq 1$ yielding $\mbox{sign}(\partial_{ii}^2 U_i) = \mbox{sign}(\partial_{ii}^2 U_i^{s}) = \mbox{sign}(\partial_{ii}^2 U_i^{d})$. That is, the sufficient condition is always met for any $w_i^s$ in $(0,1)$, if the utility of an \ac{OP} is concave with respect to $\beta_i$ for any $w_i^s$ in $[0,1]$.

\subsection{Proof of Proposition~\ref{lemma:JC}\label{ref:lemma:JC}}
The upper bound, $|\bar{J}_{ij}^{BR}|$, can be obtained by $|{\int_0^{\infty}} \frac{\mathcal{P}^s\left({\boldsymbol{\beta} \gamma}- 2{\eta} \right)\gamma}{1+\gamma}\mbox{d}\gamma| \!\!\!\overset{(p1)}{<}\!\!\! |{\int_0^{\infty}} \frac{e^{-\frac{\gamma {\boldsymbol{\beta}}}{\eta}}\left({\boldsymbol{\beta} \gamma}- 2{\eta} \right)\gamma}{1+\gamma} \mbox{d}\gamma|$ where (p1) holds if $q \!> 0$. $\bar{\kappa}_i^{(t)}$ is obtained by the RHS of (p1) for $|\bar{J}_{ij}^{BR}|$.


\end{document}